\documentclass[twocolumn,10pt]{confpaper}

\usepackage{etex}
\usepackage{comment}
\usepackage{amsmath}
\usepackage{cite}
\usepackage{url}
\usepackage{color, colortbl}
\usepackage{array}
\usepackage{algorithmic}
\usepackage{caption}
\usepackage{float}
\usepackage{graphicx}
\usepackage{subcaption}
\usepackage{color}
\usepackage{times}
\usepackage{balance}

\usepackage{mathtools}
\usepackage[font=small,format=plain,labelfont=bf,textfont=it]{caption}
\usepackage[breaklinks=true,pagebackref,filecolor=black,citecolor=blue,urlcolor=black,linkcolor=blue,colorlinks,pdfpagelabels,pdfpagelayout=SinglePage]
{hyperref}

\renewcommand{\paragraph}[1]{\vspace*{0.03in}\noindent{\bf #1}}

\graphicspath{{./figs/}}

\newcommand{\system}{{\em AlwaysOn}}

\newenvironment{itemize-s}%
  {\begin{itemize}%
    \setlength{\itemsep}{0pt}%
    \setlength{\parskip}{0pt}}%
  {\end{itemize}}

\definecolor{Gray}{gray}{0.9}

\date{}

\title{\ttlfnt{ Characterizing and Improving the Reliability of \\Broadband Internet Access}}

\author{
Zachary S. Bischof$^\dagger$\hspace{15pt} Fabi\'an E. Bustamante$^\dagger$\hspace{15pt} Nick Feamster$^\star$
\\
$^\dagger$Northwestern University
\hspace{10pt}$^\star$Princeton University
}

\begin{document}

\maketitle


\begin{sloppypar}

\begin{abstract}
In this paper, we empirically demonstrate the growing importance of
reliability by measuring its effect on user behavior. We present
an approach for broadband reliability characterization using data collected
by many emerging national initiatives to study broadband and apply it to
the data gathered by the Federal Communications Commission's Measuring
Broadband America project. Motivated by our findings, we present the design,
implementation, and evaluation of a practical approach for improving the reliability 
of broadband Internet access with multihoming.
\end{abstract}

\section{Introduction}
\label{sec:intro}

Broadband availability and performance continue to improve rapidly, spurred
by both government and private investment~\cite{itu:stateOfBb:2014} and
motivated by the recognized social and economic benefits of
connectivity~\cite{worldbank:study,itu:economic}.  The latest ITU ``State of
Broadband'' reports that there are over 60 countries where fixed or
mobile broadband penetration is above 25\%  and more than 70 countries where
the majority of the population is online~\cite{itu:stateOfBb}.  According to
Akamai's ``State of the Internet'' report, over the last four years, the top
four countries in terms of average connection speed (South Korea, Hong Kong,
Romania, and Japan) have nearly doubled their capacity~\cite{akamai:state-q3-15}.




Although providing access and sufficient capacity 
remains a challenge in many parts of the world~\cite{isaacman:c-link,zheleva:zambia}, 
in most developed countries, broadband providers are offering sufficiently high
capacities (i.e., above 10~Mbps~\cite{akamai:state-q3-15}) to encourage
consumers to migrate services for entertainment, communication and home
monitoring to over-the-top (OTT) alternatives. According to a recent survey,
nearly 78\% of U.S. broadband households subscribe to an OTT video
service~\cite{pwc:outlook16}. Enterprises are following the same path, with over 
one-third opting to use VoIP phones instead of landline ones~\cite{frost-sullivan:voip}.

%

%

\begin{figure}[t]
    \centering
\includegraphics[width=0.72\linewidth]{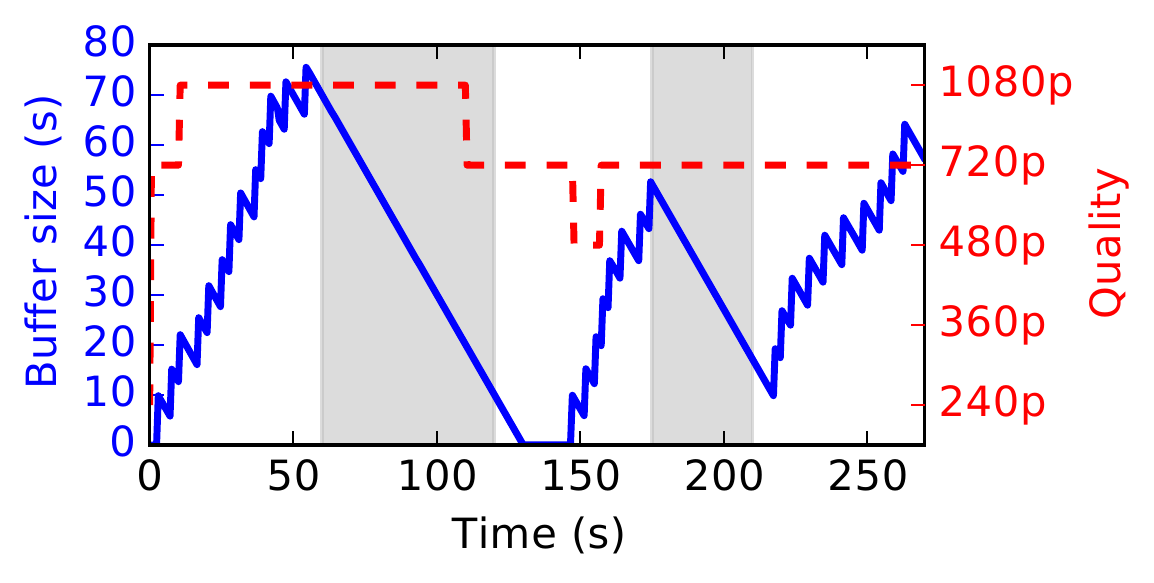}
\vspace{-10pt}
\caption{The effect of service interruption on video Quality of Experience. Even
small
outages (highlighted in gray) can have clear detrimental effects on video quality, with 
buffers draining (on the left) to the point of service interruption and the service dropping 
to a lower quality stream (on the right) in response.} \label{fig:hls_outage}
\vspace{-8pt}
\end{figure}

The proliferation of high-capacity access and the migration to OTT
services have raised users' expectations of {\em service reliability}. 
A recent survey on consumer experience by the UK Office of Communication
(Ofcom) ranks reliability first---{\em higher than even the speed of connection}---as
the main reason for customer complaints~\cite{ofcom:uk15}. 
Figure~\ref{fig:hls_outage} illustrates the impact that low reliability can
have on video Quality of Experience in a lab experiment using HTTP Live
Streaming~\cite{hls}. The plot shows the draining of the buffer - in blue -
during two service interruptions (gray bars) and the drop on video quality
(right axis) as a result.  While the buffer is quick to refill after the first
outage, the draining of it causes a drop in quality. Our empirical study of
access-ISP outages and user demand corroborates these observations, 
showing the effects of low reliability on user behavior, as captured by their demand on the network 
(\S\ref{sec:qoe}). Researchers and regulators alike have also recognized the need for clear standards 
and a better understanding of the role that service reliability plays in shaping 
the behavior of broadband users~\cite{lehr:broadband-reliability, bischof:sigcomm14:poster,
fcc:noi-11-60}. Despite its growing importance, both the reliability of 
broadband services and potential ways  to improve on it have received 
scant attention from the research community. 


In this paper, we introduce an approach for characterizing broadband
reliability using data collected by the many emerging national efforts to study
broadband (in over 30 countries~\cite{samknows:regulators})
and apply this approach to the data gathered by the Measuring Broadband
America (MBA) project, which is operated by the United States Federal Communications
Commission (FCC)~\cite{fcc:measuring}.
Motivated by our findings, we
present the design and evaluation of a practical approach to improve broadband
reliability through multihoming using a prototype implementation built as an
extension to residential gateways and a cloud-supported proxy.    

We make the following contributions: 
\begin{itemize-s}    
\item We demonstrate that poor reliability can affect user traffic demand well beyond
periods of unavailability.  For instance, we find that frequent periods of high packet loss 
(above 1\%) can result in a decrease in traffic volume for 58\% of users \textit{even during
periods of no packet loss} (\S\ref{sec:qoe}). 


    
\item We present an approach to characterize broadband service reliability. We
     apply this approach to data collected from 7,998 residential gateways over
     four years (beginning in 2011) as part of the US FCC MBA
     deployment~\cite{fcc:measuring}. We show, among other findings, that
     current broadband services deliver an average availability of at most two
     nines (99\%), with an average annual downtime of 17.8 hours
     (\S\ref{sec:characterizing}). 
     


 \item Using the FCC MBA dataset and measurements collected by over 6,000
end-host vantage points in 75 countries~\cite{otto:imc12:dnscdn}, we show that
multihoming the access link at the home gateway with two different
providers adds two nines of service availability, matching the minimum
four nines (99.99\%) required by the FCC for the public switched telephone
network (PSTN)~\cite{kuhn:pstn} (\S\ref{sec:multihoming}).  

\item We present the design, implementation, and evaluation of {\em AlwaysOn}, a 
readily deployable system for multihoming broadband connections (\S\ref{subsec:multihoming_system}).

\end{itemize-s}

\noindent  To encourage both reproducibility and use of the system, we have
publicly released our dataset, analysis scripts, and AlwaysOn
prototype~\cite{alwayson}.

\section{Importance of Reliability}
\label{sec:qoe}

In this section, we motivate the importance of reliability by studying the relationship
of service reliability and user behavior, as measured by fluctuations in user traffic
demand over time.
Studying this relationship at scale requires us to design several {\em natural
experiments}~\cite{dunning2012natural} using a subset of the data collected by
the FCC MBA effort~\cite{fcc:measuring}. We begin by  describing this dataset
and our experimental methods followed by a discussion of the results. 


\subsection{Dataset}
\label{subsec:dataset}




Since 2011, the FCC has been conducting broadband service studies using data
collected by custom gateways in selected users' homes. The collected data
includes a rich set of metrics, such as the bytes transferred per unit time, as
well as the loading times of popular websites.\footnote{A full description of
all the tests performed and data collected is available in the FCC's measuring
broadband technical appendix~\cite{fcc:2013appendix}.} This data has been
primarily used to create periodic reports on the state of broadband services in
the US as part of the MBA initiative~\cite{fcc:measuring}.  For this analysis,
we use two sets of measurements from this dataset: UDP pings and traffic byte
counters, for the FCC reports from 2011 through August 2015 (all of the
publicly available data). 

UDP pings continually measure round-trip time to at least two (typically three)
measurement servers hosted by either M-Lab or the ISPs. Every hour, the gateway
sends probes to each server at regular intervals, fewer if the link is under
heavy use for part of the hour.  The typical number of probes sent per hour
changed from 600 per hour in 2011 to approximately $2,000$ in mid-2015. 

For each target server, the gateway reports hourly statistical summaries of
latency measurements, the total number of probes sent and the number of missing
responses (i.e., lost packets). We use the latter two values to calculate the
average packet loss rate over the course of that hour. Since measurements
target multiple target servers throughout each entire hour, we select the
server with the lowest loss rate each hour for our analysis.  This prevents a
single failed target server from increasing the calculated loss rate.  

The traffic byte counters record the total number of bytes sent and received
across the WAN interface over each previous hour.  They also record counters
for the amount of traffic due to the gateway's active measurements, which we
subtract from the total volume of traffic.


\subsection{Method}
\label{subsec:methodology}

Understanding how service outages affect user behavior requires us to accurately
assess user experience in a way that is quantifiable, meaningful, and
applicable across the thousands of subscribers in the FCC dataset. For this study 
we leverage {\em natural experiments}, a class of experimental design 
common in epidemiology, the social sciences, and economics~\cite{dunning2012natural}. 

An alternative approach to explore the relationship between reliability
and user
behavior would involve controlled experiments with random
treatment---randomly assigning participants to a diverse set of connections and measuring
differences in user behavior and their experiences.
Conducting controlled experiment at scale such as this is impractical for a number
of reasons.  First, subjecting users to a controlled setting may cause deviations
from normal behavior.  Second, we want to monitor users' network use at home
under typical conditions, but doing so would require us to have control of the
quality of their access link and the reliability of their broadband service.
Intentionally degrading users' connections at the gateway would require access
to their routers (which we do not have) and would also likely deter users from
participating; even if we had access to users' home routers, changing
network conditions without their knowledge introduces complicated ethical questions.

Using natural experiments, rather than control the application of a treatment to the user, we 
test our hypotheses by measuring how users react to network conditions that occur {\em spontaneously} 
while controlling for confounding factors (e.g., service capacity, location) in our 
analysis~\cite{sitaraman:qed, bischof:imc14}. For example, to test whether high
loss rates result in decreased user demand, we compare the demand of users with
low average packet loss (our {\em control} set) to the demand of users of otherwise similar
services with high average packet loss (our {\em treatment} set).  To accept the hypothesis, 
application of the treatment should result in significantly lower network utilization. On the other hand, 
if user demand and reliability are not related, we expect the number of cases where our 
hypothesis holds to be about 50\% (i.e., random).

We use network demand as a measurable metric that may reflect user experience. Recent 
work~\cite{bischof:imc14, balachandran:video} suggests that this metric acts as
a suitable proxy for user experience. A significant change in network usage (e.g.,
bytes transferred 
or received) can be interpreted as a response to a change in the user's
experience. 

We use a one-tailed binomial test, which quantifies the
significance of deviations from the expected random distribution, to check the
validity of each hypothesis.  We consider a p-value less than 0.05 to be a
reasonable presumption against the null hypothesis ($H_{0}$).  To control for
the effects of large datasets on this class of studies, potentially making even
minor deviations significant, we only consider deviations larger than 2\% to be
practically important~\cite{paxson:strategies}. 

\subsection{Experiment results}
\label{subsec:experiment}

Several possible experiments can shed light on how service reliability affects user
behavior.  Although we expect that usage will drop around a single outage, we aim to 
understand how poor reliability over longer periods of time affects user behavior. Our
experiments test the effects on user demand of connections that are consistently 
lossy and connections that have frequent periods of high loss. 

\begin{table}
\scriptsize
\centering
\begin{tabular}{| c | c | c |}
\hline
Treatment group       & \% $H$ holds     & p-value \\ \hline
(0.5\%,    1\%)      & 48.1             & 0.792   \\
(1\%, 2\%)      & 57.7             & 0.0356  \\
$>2\%$         & 60.4             & 0.00862 \\
\hline
\end{tabular}
\vspace{-4pt}
\caption{Percentage of the time that a higher average packet loss rates will
result in lower usage. Users in the control group have similar download capacities
with an average packet loss rate between 0\% and 0.0625\%.}
\label{table:avg_pkt_loss_experiment}
\vspace{-8pt}
\end{table}


\paragraph{High average loss.} To understand how consistently lossy links affect
user demand, we calculate the average packet loss rate over the entire period
during which the user is reporting data. We then group users based on their
average packet loss rate. We select users from each treatment group and match\footnote{In 
observational studies, matching tries to identify subsamples of the treated and control 
units that are ``balanced'' with respect to observed covariates.}
them with users in the same region with similar download and upload link
capacities (within 10\% of each other) in the control group. Users in the 
control group have an average loss rate of less than 0.0625\%.  Our hypothesis, 
$H$, is that higher average packet loss rates will result in lower usage, due to a 
consistently worse experience. Our null hypothesis is that average packet loss and user
demand are not related.  Table~\ref{table:avg_pkt_loss_experiment} shows the
results of this experiment.

The results show that usage is significantly affected even for average packet losses
above 1\% --- 57.7\% of our cases show a lower volume of traffic 
with a p-value of 0.0356. This leads us to reject the null hypothesis.


\begin{table}
\scriptsize
\centering
\begin{tabular}{| c | c | c | c |}
\hline
Control         & Treatment         & \% $H$ holds  & p-value               \\ 
group           & group             &               &                       \\ \hline
(0.5\%, 1\%)    & (1\%, 10\%)       & 54.2          & 0.00143               \\
(0.1\%, 0.5\%)  & (1\%, 10\%)       & 53.2          & 0.0143                \\
(0\%, 0.1\%)    & (1\%, 10\%)       & 54.8          & 0.000421              \\
(1\%, 10\%)     & $>10\%$          & 68.3          & $3.65\times10^{-05}$  \\
(0.5\%, 1\%)    & $>10\%$         & 70.0          & $6.95\times10^{-06}$  \\
(0.1\%, 0.5\%)  & $>10\%$          & 70.8          & $2.87\times10^{-06}$  \\
(0\%, 0.1\%)    & $>10\%$          & 72.5          & $4.34\times10^{-07}$  \\ \hline
\end{tabular}
\vspace{-4pt}
\caption{Percentage of the time that users with more frequent high-loss hours
($\geq5\%$ packet loss) have lower network usage.}
\label{fig:lossy_hour_frequency_experiment}
\vspace{-8pt}
\end{table}



%

This experiment shows that a consistently lossy connection -- one with high
average packet loss -- can affect user demand. However, it is unclear if this
is caused by a change in user demand or the result of protocols reacting to
lost packets (e.g., transfer rates decreasing after a packet is dropped or
switching to lower quality streams). We attempt to address this with our next
experiment.



\paragraph{Frequent periods of high loss.} In this experiment, we test if more
frequent periods of high packet loss affects the traffic demands of users
during hours of no loss.

To understand the effects of frequent periods of high loss on user behavior we
calculate, for each user, the fraction of hours where the gateway measured more
than 5\% packet loss. We group users based on how frequently periods of high
loss occurred. For example, users that recorded loss rates above 5\% during 0\%
to 0.1\% of measurements were placed in a group that we used as one of the
controls.  We then compared the network demands during peak hours with no
packet loss between each pair of user groups. In this case, our hypothesis,
$H$, is that groups with a high frequency of high loss rates (treatment group)
will have lower usage than groups with a low frequency of high loss rates
(control group). Table~\ref{fig:lossy_hour_frequency_experiment} shows the
results of this experiment. 

We find that users with high packet loss rates during more than 1\% of hours,
tend to have lower demand on the network during periods of no packet loss.  As
the difference between the frequency of high loss rates periods increases, the
magnitude of this effect increases, with larger deviations from the expected
random distribution.

Previous studies have discussed the importance of broadband service
reliability~\cite{lehr:broadband-reliability}, and surveys of broadband users
have shown that reliability, rather than performance, has become the main
source of user complaints~\cite{ofcom:uk15}; our findings are the first to
empirically demonstrate the relationship between service reliability and user
traffic demand.

\section{Characterizing Reliability}
\label{sec:characterizing}

We now present an approach for characterizing broadband service reliability
that can apply to the datasets that many ongoing national broadband measurement
studies are collecting.   The decision to make our analysis applicable to the
existing national datasets introduces several constraints on our analysis
method, including the type and granularity of metrics and the placement of
vantage points.
At the same time, our approach is applicable to the various available datasets.
Ultimately, our work can motivate future experiment designs to better capture
all aspects of broadband service reliability.  We describe ongoing broadband
measurement efforts before presenting our methods and metrics for
characterizing service reliability.  We then discuss our findings concerning
the reliability of broadband services in the US.

\subsection{Approach}
\label{subsec:approach}


\paragraph{Available data.}
Over the last decade, the number of governments with national broadband plans
has increased rapidly~\cite{itu:nbp}, and several of these governments
are funding studies to characterize the broadband services
available to their citizens.  Two prominent examples are the efforts being
carried out by the UK Ofcom and the US FCC in collaboration with the UK company
SamKnows.  In the few years since their initial work with Ofcom, SamKnows 
has begun working with at least six additional governments including 
the US, Canada, Brazil, the European Union and Singapore. Data for these efforts 
is typically collected from modified residential gateways distributed to
participants in a range of service providers. 

We use the FCC's dataset for our characterization of broadband reliability, as
it is the only effort that currently publishes copies of its raw data. In
addition to using the techniques in Section~\ref{subsec:dataset} to clean the
data, we also attempt to validate a user's ISP by looking at the gateway's
configured DNS IP addresses, making sure they are consistent with subscribing
to that provider (e.g., a user listed as Comcast user should be direct to a DNS
server in Comcast's network). We also remove any gateways that have been
marked for removal by the FCC's supplemntary unit metadata.

\paragraph{Metrics.} To analyze the data from these efforts, we use a number
of conventional metrics to quantify the reliability of broadband service.
These metrics are defined based on an understanding of what constitutes a
failure. We define the {\em reliability} of a broadband service as the average
length of time that the service is operational in between interruptions and 
{\em
availability} as the fraction of time the service is in functioning condition.


We adopt several well-accepted metrics from reliability engineering, including
{\em Mean Time Between Failure} (MTBF) and {\em Mean Down Time}
(MDT).  MTBF is the average time that a service works without failure; it is
the multiplicative inverse of Failure Rate, formally defined as  {\small \[MTBF =
\frac{\text{Total uptime}}{\text{\# of failures}}\]} 

\noindent
To characterize the length of time a service is unavailable during each
failure, we use MDT, which is defined as {\small \[MDT = \frac{\text{Total
downtime}}{\text{\# of failures}}\]}

\noindent
We can now define availability ($A$) as the probability
that at any given point in time, the service is functioning/operational.
Unavailability is the complement of availability. 
More formally {\small \[A =
\frac{\text{MTBF}}{\text{MTBF}+\text{MDT}}\]
\[U = (1 - A).\]}

\paragraph{Definition of a failure.} What constitutes a failure or outage in
the context of broadband services is a critical issue tightly coupled to the
collected metrics. Although the definition of failure is obvious in many systems, it is
less clear in the context of ``best-effort'' networks.
 
\begin{figure}[t]
    \centering
    \includegraphics[width=.8\linewidth]{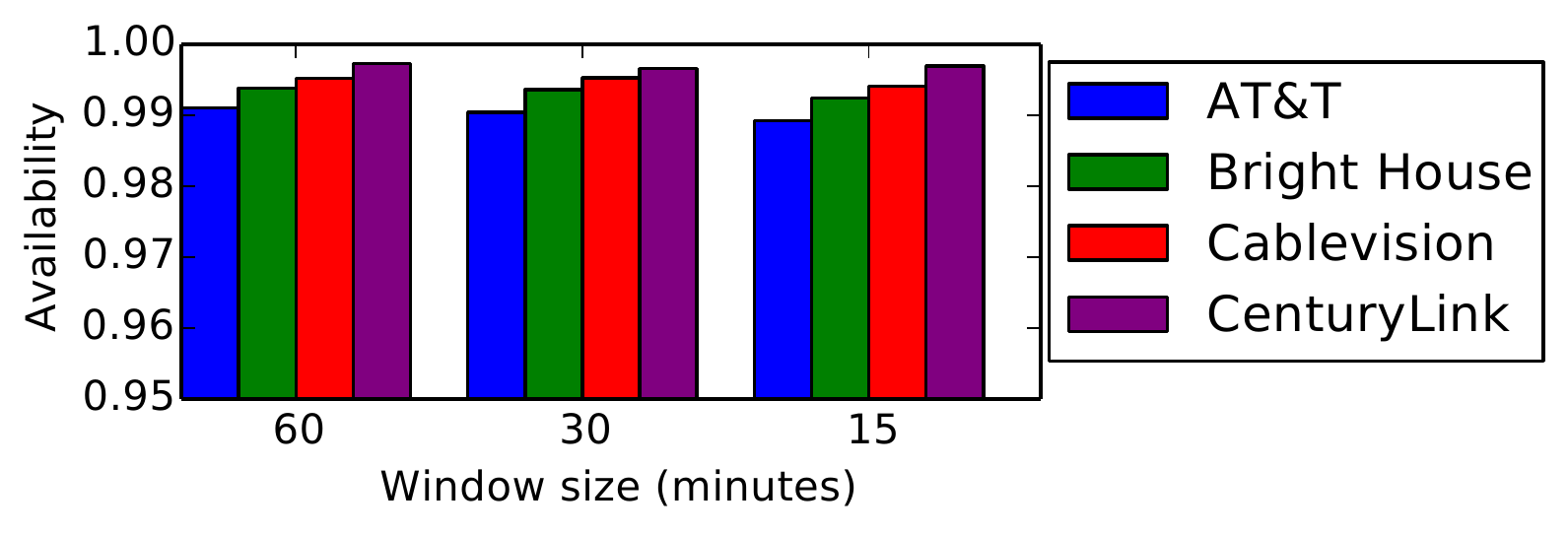}
    \vspace{-8pt}
    \caption{Service availability for four ISPs across multiple 
    observation window sizes.
    \label{fig:loss_window}}
    \vspace{-8pt}
\end{figure}

\begin{figure}[t]
    \centering
    \includegraphics[width=.75\linewidth]{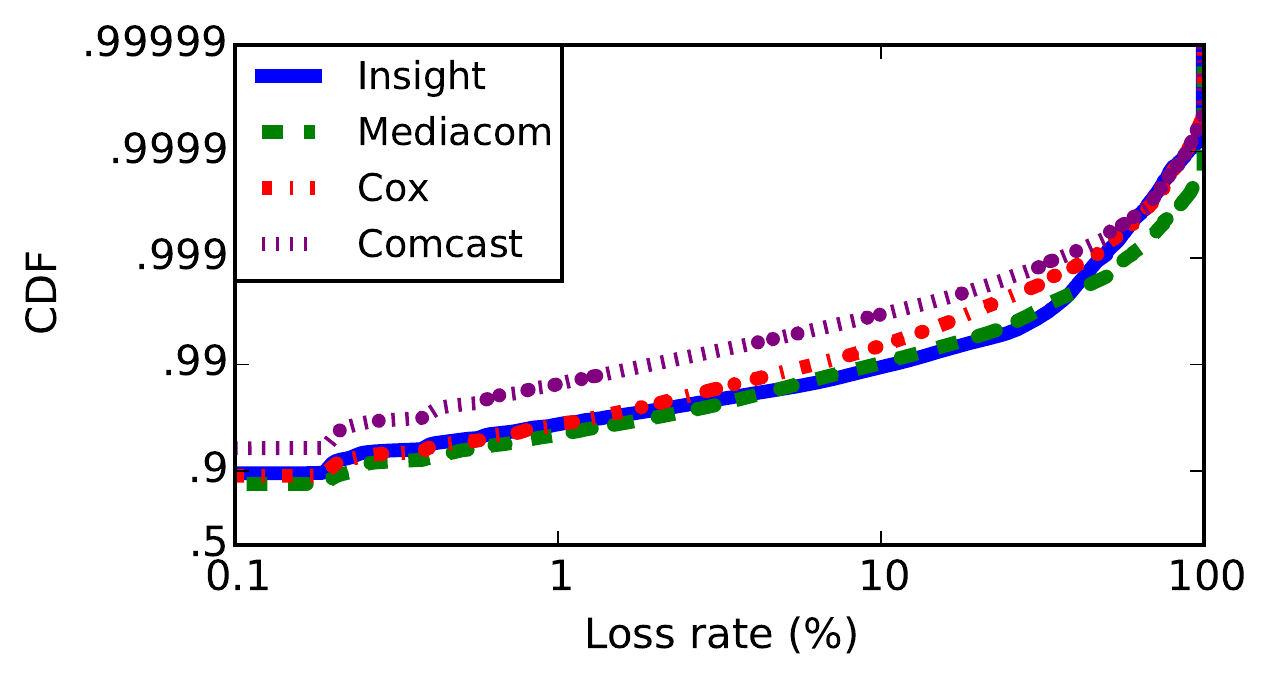}
    \vspace{-8pt}
    \caption{Hourly loss rates measured from gateways of 
    four cable providers. Lower curves indicate a less available service; curves 
    crossing over each other implies that different loss-rate thresholds would 
    yield different rankings.}
    \label{fig:cable_loss_ccdf}
    \vspace{-8pt}
\end{figure}

We choose to identify connections failures by detecting significant changes in
lost packets. It is unclear what packet loss rate (or rates) should be used as
thresholds for labeling failures. Achievable TCP throughput varies inversely
with the square root of loss rate~\cite{mathis:tcp,padhye:tcp} and even modest
loss rates can significantly degrade performance. Xu et al.  showed that video
telephony applications can become unstable at a 2\% bursty
loss~\cite{xu:imc12:video}, with significant quality degradation occurring
around 4\% in some cases. In our analysis, we use three thresholds for
classifying network failures -- 1\%, 5\%, and 10\%.

While the FCC MBA dataset is currently the largest publicly available dataset 
on broadband service performance, relying on it for our analysis means we 
are only able to measure loss rates at a one-hour granularity. To evaluate the
impact of monitoring granularity,  we rely on a platform installed in 6,000 end-hosts
to measure loss by sending packets approximately every five seconds and 
use this data to calculate loss rate using different sizes and loss rate thresholds. 
Figure~\ref{fig:loss_window} shows the availability of four ISPs in our dataset using 
the 10\% loss threshold. We found that changing the window size has
little impact on our calculation of availability and the relative ranking of ISPs.

The distribution of loss rates are quite different for different broadband
technologies, and can vary even across providers with the same technology at
different loss rate thresholds. Figure~\ref{fig:cable_loss_ccdf} shows the CDF
of loss rate of four cable providers, with the y-axis showing the cumulative
fraction of all hourly time intervals. Although two providers may offer the
same MTBF for a particular loss rate threshold, considering the difference in
loss rate distributions, a different definition of ``failure'' could result in
a different ranking. For instance, defining a failure as ``an hour with $>1\%$
packet loss'' yields a similar MTBF for both Cox and Insight Cable
($\approx~27.5$ hours), using a 10\% loss rate threshold, but results in a MTBF
over 50\% higher for Cox ($\approx~150$ hours) than for Insight ($\approx~94$
hours). 

The assessment of broadband reliability could focus on different aspects,
ranging from the reliability of the network connection, the consistency of
performance, and the availability of services offered by the ISP, such as DNS
servers and email~\cite{lehr:broadband-reliability}.  The primary focus of this
work is on broadband service reliability, under which we include both the
availability of connection itself as well as that of the ISP's DNS service.
From the perspective of most users, failures in either are indistinguishable.
We plan to study other aspects of service reliability, such as performance
consistency, in future work.

\subsection{Characterization of service reliability}

We apply the approach presented in the previous section to
characterize the reliability of broadband services in the US using the 
FCC MBA dataset.
We first provide a short summary of the population of participants in the
SamKnows/FCC study. In our study, we seek to understand the role
that a set of key attributes of a subscriber's connection play in determining
its reliability: (1)~How does reliability vary across different providers?
(2)~What is the impact of using different access technologies or
subscribing to different tiers of service? (3)~Does geography affect reliability?
(4)~How reliable is the provider's DNS service?

\begin{table}[t]
\scriptsize
\centering
\begin{tabular}{| c | c |}
\hline
Technology  &   \% of participants \\ \hline
Cable       &   55\% \\
Cable (business) & 1\% \\
DSL         &   35\% \\
Fiber       &   7\% \\
Satellite   &   1\% \\
Wireless    &   1\% \\ \hline
\end{tabular}
\vspace{-8pt}
\caption{Percentage of the sample population in the FCC's dataset using each
access link technology.\label{fig:fcc_technologies_pie}}
\vspace{-12pt}
\end{table}

\paragraph{Sample population.} As part of the MBA
dataset, the FCC also provides metadata about each participant
including the user's service tier (i.e., subscription speed), service
technology (e.g., cable or DSL), and geographic location.  Combining this
information with the loss rate data described in Section~\ref{subsec:approach},
we compare the reliability of broadband services across different axis. 

The list of ISPs covered in the sample population includes both large,
nationwide ISPs and smaller, regional ISPs. Since the number
of devices per ISP is weighted by the number of subscribers, most devices
(71\%) are located in larger ISPs (AT\&T, Comcast, and Verizon).


The FCC's dataset includes a diverse set of technologies, including satellite
and fixed wireless providers.  Table~\ref{fig:fcc_technologies_pie} shows a
summary of the distribution of participants by access technology. 
``Wireless'' access refers to fixed wireless (not mobile) from providers such
as Clearwire, where users connected there FCC-provided device to a wireless
modem.  Additional information, such as the process used for selecting
participants, can be found in the technical appendix of the FCC's
report~\cite{fcc:2013appendix}. 

To understand the relative importance of the different attributes, we 
calculated the information gain---the degree to which a feature is 
able to reduce the entropy of a target variable---of each attribute of a
subscriber's
connection (ISP, download/upload capacity, region, and access technology). We
found
the subscriber's ISP to be the most informative feature, with access link
technology
as a close second, for predicting service availability. In the rest of this
section
we analyze the impact of these attributes on service reliability. We close with
an
analysis of DNS and ISP reliability.

\subsubsection{Effect of ISP}
\label{subsubsec:char-isp}

\begin{table}[t]
\scriptsize
\centering
\begin{tabular}{| l | c | c | c | r | r | r |}
\hline
ISP             & \multicolumn{3}{c |}{Average}      & \multicolumn{3}{ c
|}{Average annual} \\
& \multicolumn{3}{c |}{availability} & \multicolumn{3}{ c |}{downtime (hours)}
\\ \hline
&   1\% &   5\% &   10\%            & 1\%   &   5\%   &   10\% \\ \hline
\multicolumn{7}{| l |}{\textit{Fiber}}  \\ \hline
\hspace{4pt}
Frontier (Fiber)& 98.58 & 99.47 & 99.77             & 124   & 46.8  & 20.3 \\
\rowcolor{Gray}
\hspace{4pt}
Verizon (Fiber) & \textbf{99.18} & 99.67 & 99.80             & 72  & 29.2  & 17.8 \\
\hline
\multicolumn{7}{| l |}{\textit{Cable}}  \\ \hline
\hspace{4pt}
Bright House    & 98.21 & 99.28 & 99.58             & 156   & 62.8  & 36.7 \\ 
\hspace{4pt}
Cablevision     & 98.33 & 99.53 & 99.70             & 146   & 41.4  & 25.9 \\ 
\hspace{4pt}
Charter         & 97.84 & 99.29 & 99.59             & 189   & 62.5  & 36.1 \\
\hspace{4pt}
Comcast         & 98.48 & 99.45 & 99.66             & 134   & 48.0  & 29.7 \\
\hspace{4pt}
Cox             & 96.35 & 98.82 & 99.33             & 320   & 103.0   & 58.4 \\
\rowcolor{Gray}
\hspace{4pt}
Insight         & 96.38 & 98.31 & \textbf{98.94}             & 318   & 148.0   & 93.0 \\
\hspace{4pt}
Mediacom        & 95.48 & 98.34 & 99.03             & 396   & 146.0   & 85.3 \\
\hspace{4pt}
TimeWarner      & 98.47 & 99.48 & 99.69             & 134   & 45.9  & 26.9 \\
\hline
\multicolumn{7}{| l |}{\textit{DSL}}  \\ \hline
\hspace{4pt}
AT\&T           & 96.87 & 99.05 & 99.42             & 274   & 83.3  & 51.1
\\
\hspace{4pt}
CenturyLink     & 96.33 & 98.96 & 99.39             & 322   & 90.9  & 53.7 \\
\rowcolor{Gray}
\hspace{4pt}
Frontier (DSL)  & 93.69 & 98.18 & \textbf{98.87}             & 553   & 160.0   & 98.7 \\
\hspace{4pt}
Qwest           & 98.24 & 99.24 & 99.51             & 154   & 66.7  & 42.8 \\
\hspace{4pt}
Verizon (DSL)   & 95.56 & 98.43 & 99.00             & 389   & 137.0   & 88.0 \\
\hspace{4pt}
Windstream      & 94.35 & 98.72 & 99.42             & 495   & 112.0   & 50.6 \\
\hline
\multicolumn{7}{| l |}{\textit{Wireless}}  \\ \hline
\rowcolor{Gray}
\hspace{4pt}
Clearwire       & 88.95 & 96.96 & \textbf{98.13}             & 968   & 266.0   & 164.0
\\ \hline
\multicolumn{7}{| l |}{\textit{Satellite}}  \\ \hline
\hspace{4pt}
Hughes          & 73.16 & 90.15 & 94.84             & 2350  & 863.0   & 453 \\
\hspace{4pt}
Windblue/Viasat & 72.27 & 84.20 & 96.37             & 2430  & 1380.0  & 318.0 \\
\hline
\end{tabular}
\caption{Average availability and annual downtime for 
subscribers, per service, for three different loss-rate thresholds. Verizon (fiber)
is the only service providing two nines of availability at the 1\%
loss rate threshold. Clearwire is able to reach performance close to 
Frontier (DSL) 
and Insight at the 10\% threshold.
}
\label{table:isp_availability_downtime}
\vspace{-14pt}
\end{table}

 
We first characterize service {\em availability}--- the probability that a
service is operational at any given point in time---for each provider in our
dataset.  Table~\ref{table:isp_availability_downtime} lists the average
availability per ISP, as well as the provider's unavailability, described
as the average annual downtime (in hours).  We evaluate both metrics in the
context of the three loss rate thresholds for network failures
measured over an hour. For
comparison,
five nines is often the target availability in telephone
services~\cite{fivenines}. 

We find that, at best, some providers are able to offer two nines of
availability. Verizon's fiber service is the only one with two nines of
availability at the 1\% threshold. At 5\%, about half of the providers offer
just over two nines.  The satellite and wireless services from Clearwire,
Hughes, and Viasat provide only one nine of availability, even at the 10\% loss
rate threshold. 

\begin{table}[t]
\scriptsize
\centering
\begin{tabular}{| l | r | r | r | r |}
\hline
ISP & $A$ & \% change in $U$ & $A$ & \% change in $U$ \\ \hline
    &  \multicolumn{2}{c |}{1\%} & \multicolumn{2}{ c |}{10\%} \\ \hline
\multicolumn{5}{| l |}{\textit{Satellite}}  \\ \hline
\hspace{4pt}
Hughes          & 60.97 & +45.4  & 91.38 & +66.9 \\
\hspace{4pt}
Wildblue/ViaSat & 69.44 & +10.2  & 94.14 & +61.2 \\ \hline
\multicolumn{5}{| l |}{\textit{Wireless}}  \\ \hline
\hspace{4pt}
Clearwire       & 86.35 & +23.6  & 97.57 & +29.9 \\ \hline
\multicolumn{5}{| l |}{\textit{DSL}}  \\ \hline
\rowcolor{Gray}
\hspace{4pt}
Windstream      & 89.17 & +91.8  & 99.13 & \textbf{+50.4} \\
\hspace{4pt}
Frontier (DSL)  & 87.98 & +90.4  & 98.42 & +39.9 \\
\hspace{4pt}
Verizon (DSL)   & 93.95 & +36.2  & 98.90 & +9.9 \\
\hspace{4pt}
CenturyLink     & 94.19 & +58.2  & 99.35 & +6.9 \\
\hspace{4pt}
AT\&T           & 95.85 & +32.4  & 99.38 & +5.4 \\
\hspace{4pt}
Qwest           & 97.92 & +18.5  & 99.51 & +1.2 \\ \hline
\multicolumn{5}{| l |}{\textit{Cable}}  \\ \hline
\rowcolor{Gray}
\hspace{4pt}
Cablevision     & 97.76 & +34.2  & 99.64 & \textbf{+22.6} \\
\hspace{4pt}
TimeWarner      & 98.03 & +28.5  & 99.69 & +1.3 \\
\hspace{4pt}
Insight         & 95.31 & +29.4  & 98.98 & -3.9 \\
\hspace{4pt}
Charter         & 97.75 & +4.2   & 99.61 & -6.4 \\
\hspace{4pt}
Mediacom        & 94.52 & +21.1  & 99.09 & -7.0 \\
\hspace{4pt}
Comcast         & 98.39 & +5.3   & 99.70 & -11.7 \\
\hspace{4pt}
Brighthouse     & 98.15 & +3.5   & 99.63 & -11.8 \\
\rowcolor{Gray}
\hspace{4pt}
Cox             & 96.30 & +1.3   & 99.42 & \textbf{-13.3} \\ \hline
\multicolumn{5}{| l |}{\textit{Fiber}}  \\ \hline
\hspace{4pt}
Frontier (Fiber) & 98.56 & +1.4   & 99.78 & -4.6 \\
\rowcolor{Gray}
\hspace{4pt}
Verizon (Fiber)   & 99.11 & +8.7   & 99.83 & \textbf{-14.7} \\ \hline
\end{tabular}
\vspace{-4pt}
\caption{Average availability ($A$) and percent change in unavailability ($U$)
    for subscribers of each ISP during peak hours. Some providers had
significantly higher unavailability at the 10\% threshold 
during peak hours, including Windstream and 
Cablevision, as well as satellite and wireless services. Cox and Verizon (fiber)
had the largest improvement in availability during peak hours, as outages 
were concentrated during early morning or mid-day.}
\label{table:isp_availability_peak}
\vspace{-8pt}
\end{table}


Because broadband users are more likely to be affected by outages in the
evening, we also measured availability during peak hours (from 7PM to 11PM,
local time), as shown in Table~\ref{table:isp_availability_peak}.  Although all
providers show a lower availability at the 1\% loss rate threshold compared to
their full-day average, most cable providers actually performed better at a
10\% loss rate threshold. We expect that some of these providers may perform
planned maintenance, which would introduce extremely high periods of loss
($>10\%$), during the early morning or midday.  Overall, Cox and Verizon
(fiber) had the largest decrease in unavailability during peak hours.

On the other hand, DSL, wireless, and satellite providers continued to have
lower availability at the 10\% threshold during peak hours, as compared to
their average availability over all time periods. Of all the cable providers,
only two had an increase in unavailability during peak hours, with Cablevision
having the biggest change. Windstream and Frontier (DSL) also had a much larger
increase in unavailability during peak hours compared to other DSL providers.

We also analyzed the MTBF for each provider, which represents the average time 
between periods with high packet loss. Most ISPs appear to maintain a MTBF of 
over 200 hours ($\approx$~8 days), but a few experience failures every 100 hours,
on average. ClearWire, Hughes, and Viasat again have notably low
MTBF: 73.8, 26.0, and 4.78 hours, respectively. CenturyLink and Mediacom offer
the two lowest MTBFs for DSL and cable providers, respectively. 
These network outages are resolved, on average, within one to two hours for
most ISPs. The main exception is satellite providers---more specifically Viasat---
with a MDT (mean downtime), close to 5.5 hours.  


%

In general, most ISPs ranked similarly across both MTBF and MDT, with a few
exceptions.  For instance, Verizon's fiber service had the highest MTBF, but
its periods of downtime were often over 2.5 hours. Frontier's DSL service, on
the other hand, had frequent failures, but these periods of failure were
relatively short.

\subsubsection{Effect of access technology}
\label{subsubsec:technologies}

\begin{figure}[t]
\begin{subfigure}{0.70\linewidth}
    \centering
    \includegraphics[width=\linewidth]{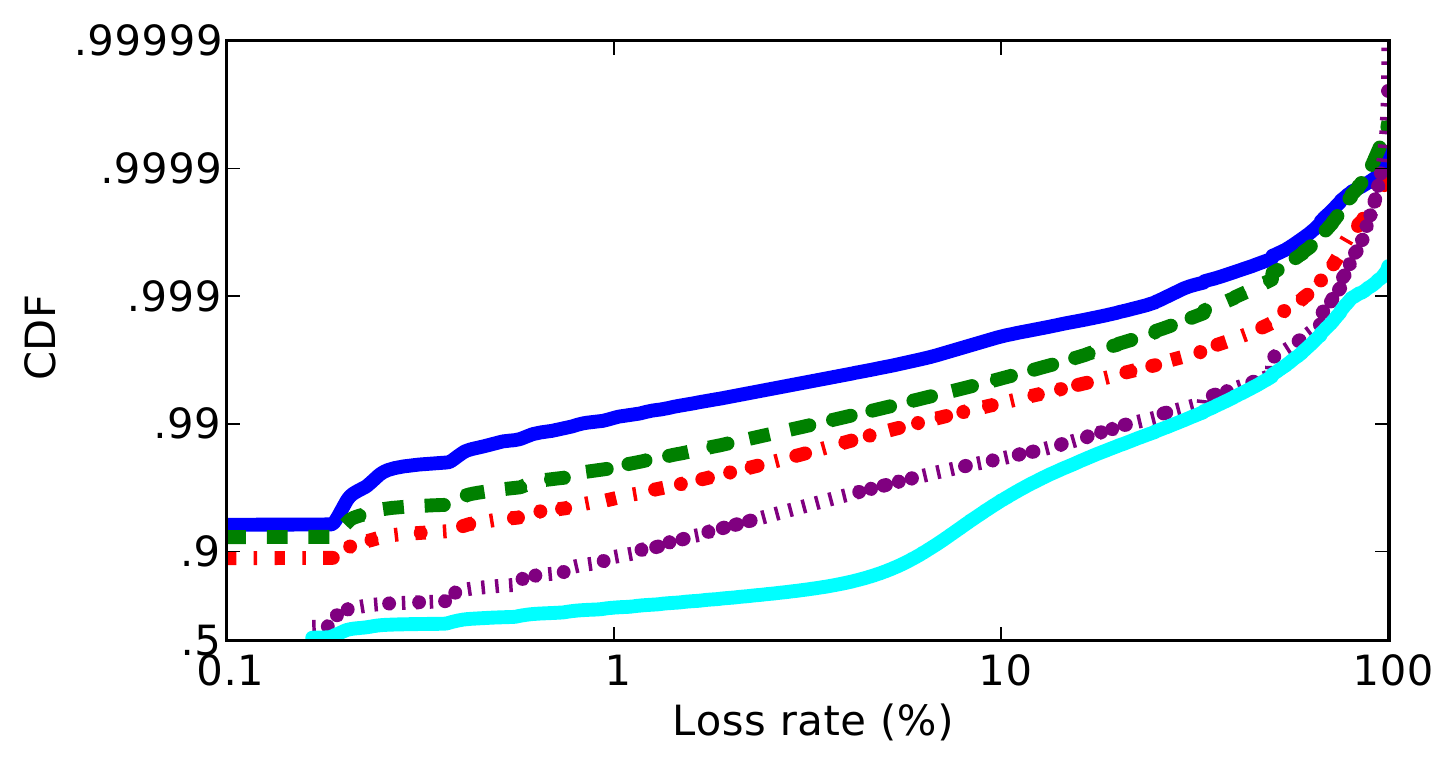}
    \caption{Hourly loss rates
    \label{fig:technology-loss-rate-ccdf}}
\end{subfigure} \begin{subfigure}{0.2\linewidth}
\includegraphics[width=\linewidth]{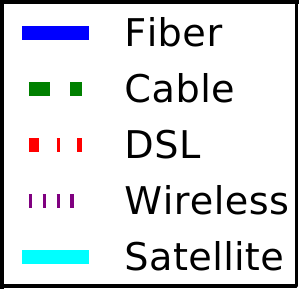}
\end{subfigure}\\
\begin{subfigure}{0.75\linewidth}
    \centering
\includegraphics[width=\linewidth]{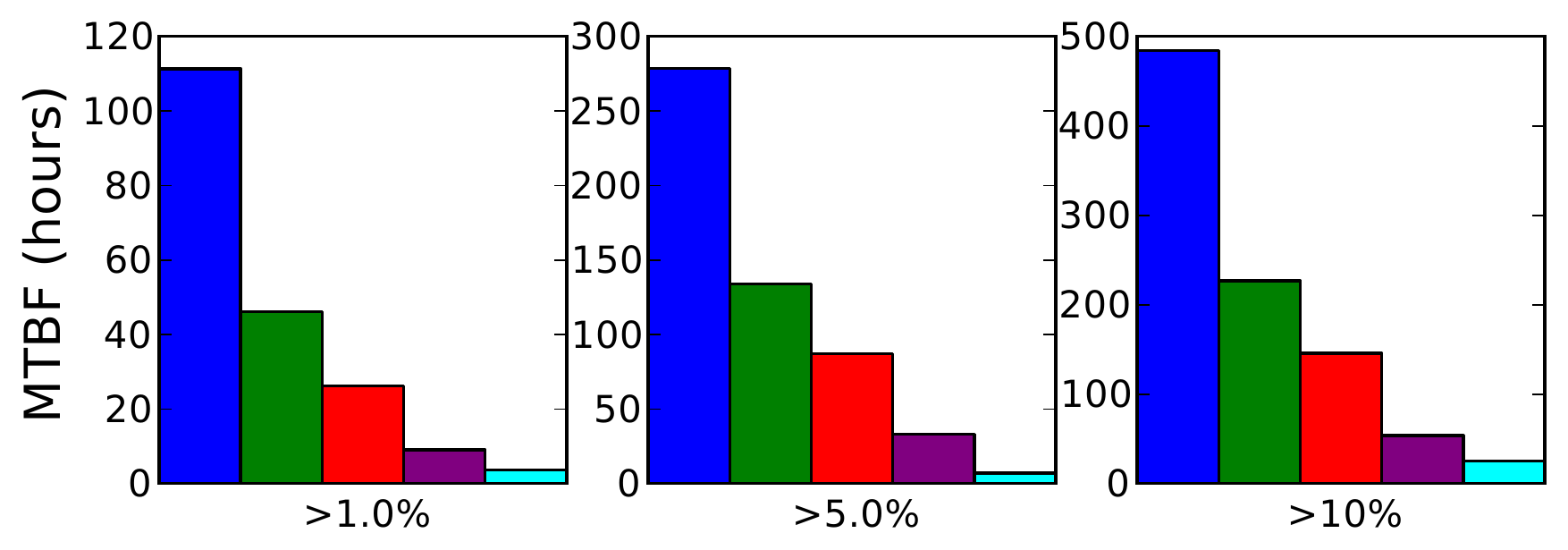}
    \caption{MTBF \label{fig:technology-loss-rate}}
\end{subfigure}
\caption{Hourly loss rates and MTBF for each type of access technology. There
is a clear separation between technology for both metrics.}
\end{figure}

Next, we study the impact of a subscriber's access technology.  
Figure~\ref{fig:technology-loss-rate-ccdf} shows a CDF of packet loss rates for
each access technology. As expected, we find that fiber services provide the
lowest loss
rates of all technologies in our dataset with only 0.21\% of hours having
packet loss
rates above 10\%. Stated differently, fiber users could expect an hour with
10\% packet loss to occur approximately once ever 20 days. Cable and DSL services
are next in terms of reliability, with periods of 10\% packet loss only
appearing 0.44\% and 0.68\% of the time, respectively.  Periods with packet loss
rates above 10\% were almost a full order of magnitude more frequent for
wireless (1.9\%) and satellite (4.0\%) services.

We compare the average interval between hours with loss above the different
loss-rate
thresholds, shown in Figure~\ref{fig:technology-loss-rate}.  For each
threshold, fiber
performs significantly better, with cable and DSL again showing relatively
similar performance. 

\begin{figure}
    \centering
\begin{subfigure}{0.45\linewidth}
    \centering
    \includegraphics[width=\linewidth]{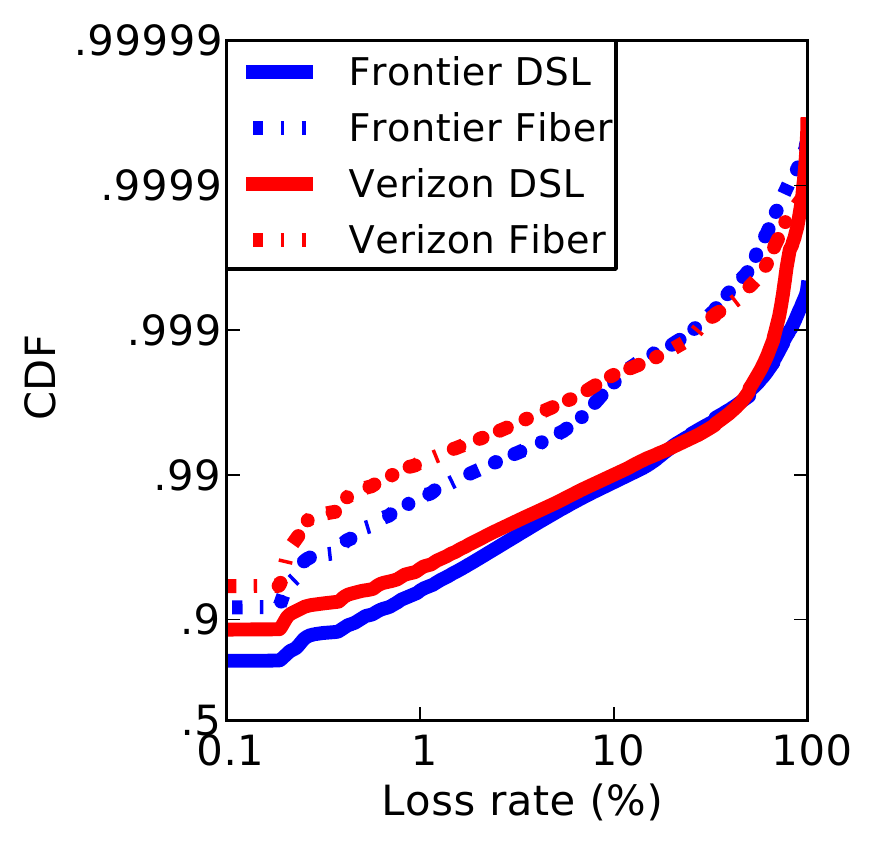}
\vspace{-12pt}
    \caption{Fiber vs. DSL.    \label{fig:technology-loss-rate-by-isp}}
\end{subfigure}
\begin{subfigure}{0.45\linewidth}
    \centering
\includegraphics[width=\linewidth]{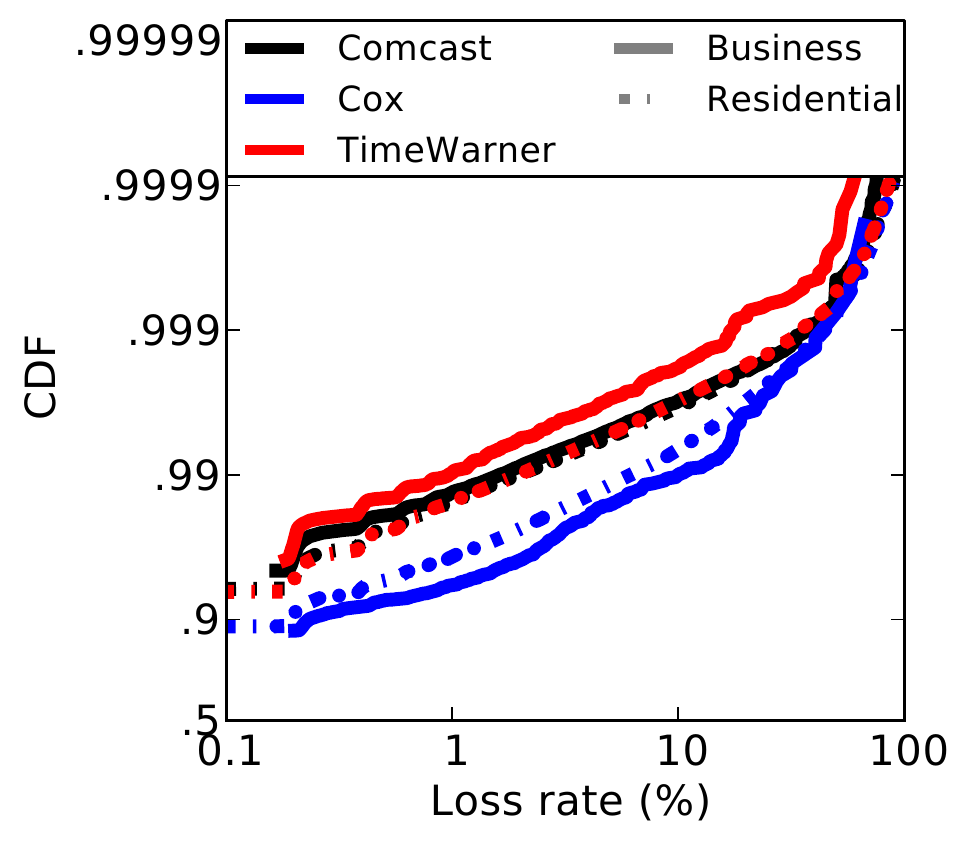}
    \caption{Business vs. residential.   \label{fig:service-class}}
\end{subfigure}
\caption{The hourly loss rates for subscribers of each service.
Technology,
rather than provider, is the main determinant of availability and service tiers
has little effect.}
\end{figure}

Other factors that affect the reliability may in fact be related to access
technology; for example, network management policies of a particular ISP might
be correlated with the ISP's access technology and could hence play a role in
determining network reliability. To isolate such effects, we compare the
difference in service reliability within the same provider, in the same
regions, but for different technologies. Only two providers offered broadband
services over more than one access technology: Frontier and Verizon, both of
which have DSL and fiber broadband services.
Figure~\ref{fig:technology-loss-rate-by-isp} shows a CDF of the loss rates
measured by users of both services. Although there are differences across the
two providers, in general, subscribers using same access technology tend to
experience similar packet loss rates. Verizon and Frontier DSL customers
measured high loss rates (above 10\%) during 1.56\% and 1.82\% of hours, while
Verizon and Frontier fiber customers saw high loss rates during 0.33\% and
0.53\% of hours.

\subsubsection{Effect of service tier}
\label{subsubsec:service_class}

In addition to offering broadband services over multiple access technologies, a
number of ISPs offer different service tiers on the same access technology. For
example, Comcast, Cox, and Time Warner all have a ``business class'' service in
addition to their standard residential service. We explored how reliability
varies across different service tier offerings within the same provider.  

Figure~\ref{fig:service-class} shows a CDF of the loss rates reported
by users of each provider's residential and business class Internet
service.  In general, the service class appeared to have little effect on the 
reliability of a service. The differences in packet loss rates are small
compared to the difference between access technologies in the same provider.
Comcast business subscribers see about the same loss rates as the residential
subscribers, while Time Warner's business subscribers report slightly
lower packet loss rates. On the other hand, Cox business subscribers actually
report a slightly higher frequency of packet loss when compared to residential
subscribers.  In particular, there are occasionally anecdotes that providers
might be encouraging subscribers to upgrade their service tier by offering
degraded service for lower service tiers in a region where they were offering
higher service tiers; we did not find evidence of this behavior.

\subsubsection{Effect of demographics}
\label{subsubsec:geographical}

We also explored the relationship between population demographics and the
reliability of Internet service.  For this we combined publicly
available data from the 2010 census with the FCC dataset to see how factors
such as the fraction of the population living in an urban setting, population
density and gross state product per capita relate to network reliability. 

We looked at service reliability and urban/rural population distributions per
state using the classification of the US Census Bureau with ``urbanized areas''
($>$ 50,000 people), ``urban clusters'' (between 2,500 and 50,000 people), and
``rural'' areas ($<$ 2,500 people)~\cite{census-bureau-def}. We also explore
the correlation between failure rate and a state's gross state product (GSP)
per capita. 

Overall, we found a weak to moderate correlation between failure rates and
both percent of urban population ($r = -0.397$) and GSP per capita ($r =
-0.358$), highlighting the importance of considering context when
comparing the reliability of service providers (the direction of the
causal relationship is an area for further study).

\subsubsection{ISP and DNS reliability}
\label{subsubsec:services}

We also include a study of ISPs' DNS service availability in our analysis of
broadband reliability. Previous work has shown that DNS plays a significant
role in determining application-level
performance~\cite{wang:wprof,otto:imc12:dnscdn} and thus users' experience.
Additionally, for most broadband users, the effect of a DNS outage is identical
to that of a faulty connection.

For DNS measurements, the gateway issues an hourly A record query to both the
primary and secondary ISP-configured DNS servers for ten popular websites. For
each hostname queried, the router reports whether the DNS query succeeded or
failed, the response time and the actual response.  Every hour, the
FCC/SamKnows gateway performs at least ten queries to the ISP-configured DNS
servers. For this analysis, we calculate the fraction of DNS queries that fail
during each hour. To ensure that we are isolating DNS availability from access
link availability, we discard hours during which the gateway recorded a loss
rate above 1\%.  This corresponds to less than 3\% of hours in our dataset.  We
classified hours where the majority of DNS queries failed (over 50\%) as
periods of DNS unavailability. 

\begin{figure}[t]
    \centering
\includegraphics[bb=7 8 332 273,
width=.7\linewidth]{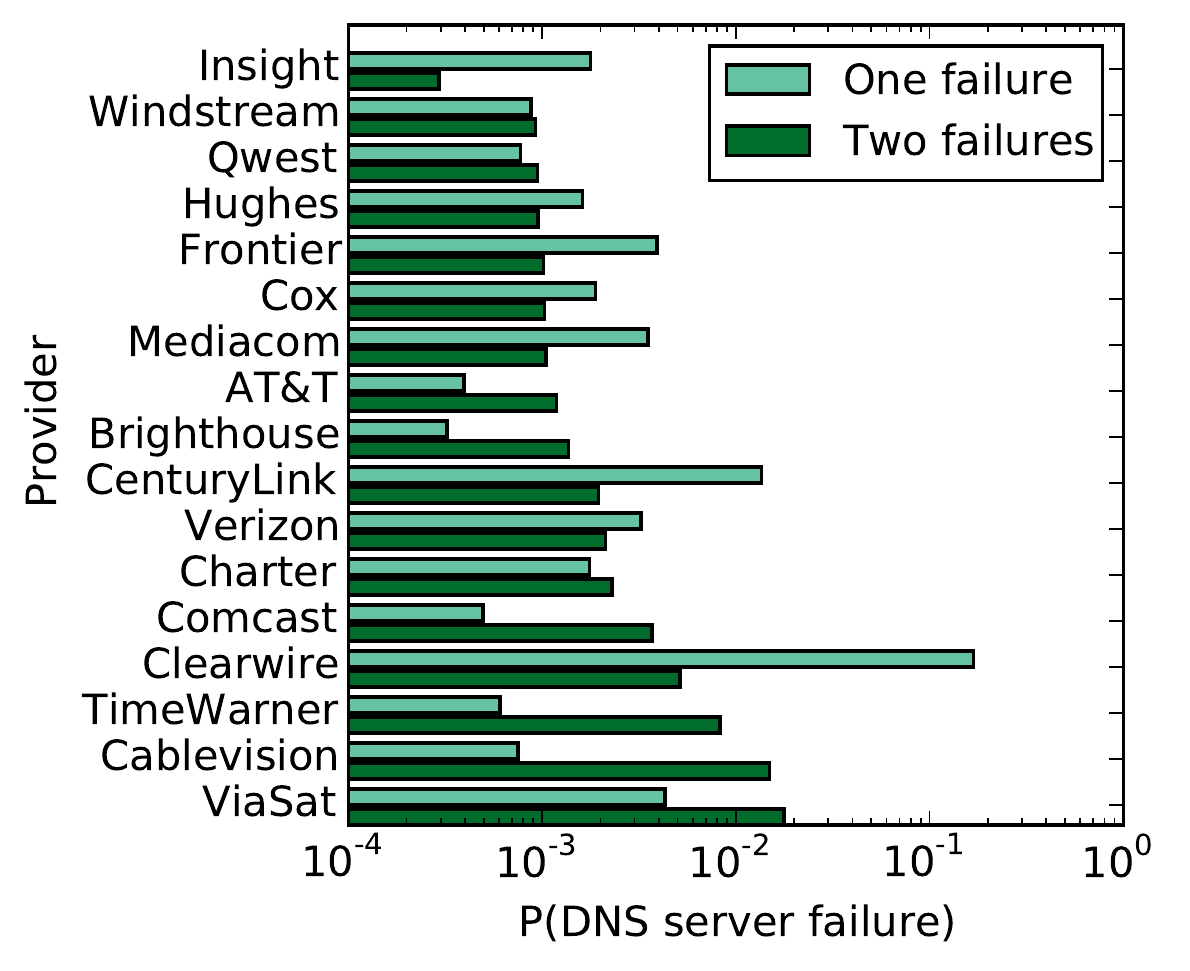}
\vspace{-4pt}
    \caption{Probability that one (or two) DNS servers will be unavailable 
    for each ISP's configured DNS servers.  We consider the two cases
    independently (i.e., ``one failure'' reflects the event that exactly
one server fails to respond to queries).}
    \label{fig:dns-failure-rate}
\vspace{-8pt}
\end{figure}


Figure~\ref{fig:dns-failure-rate} shows the probability of each provider
experiencing one and two DNS server failures during a given hour. We sort
providers in ascending order based on the probability that two servers will
fail during the same hour.  

Surprisingly, we find that many ISPs have a higher probability of two
concurrent failures than a single server failing. For example, Comcast's
primary and secondary servers are almost an order of magnitude more likely to
fail simultaneously than individually.\footnote{One possible explanation is the 
reliance on anycast DNS. We are exploring this in ongoing work.}

As one might expect, a reliable access link does not necessarily imply a
highly available DNS service. For example, in our analysis of the reliability
of access link itself, Insight was in the middle of the pack in terms of
availability, offering only one nine of availability
(Table~\ref{table:isp_availability_downtime}), yet the results in 
Figure~\ref{fig:dns-failure-rate} show Insight having the lowest probabilities 
that queries to both DNS servers would fail simultaneously.

\subsubsection{Longitudinal analysis}
\label{subsubsec:longitudinal}

\begin{figure*}
    \centering
\begin{minipage}{1\linewidth}
\begin{subfigure}{0.24\linewidth}
    \includegraphics[width=\linewidth]{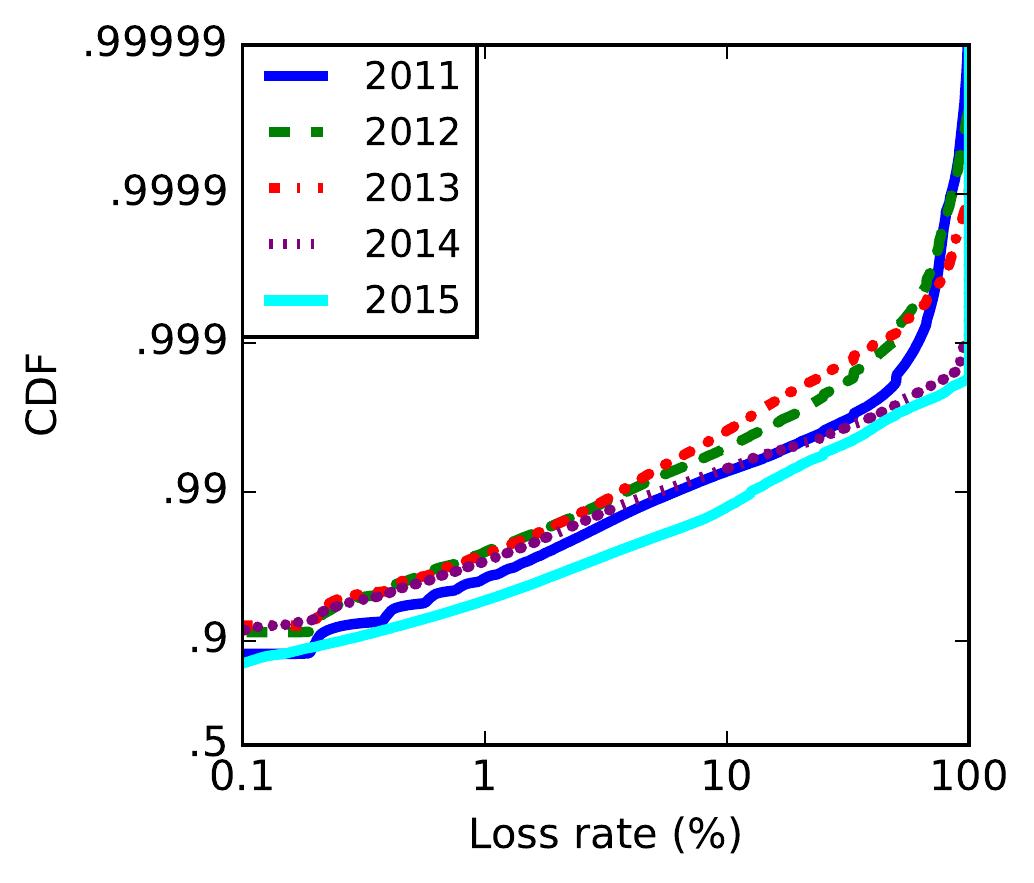}
    \caption{AT\&T\label{fig:long-att}}
\end{subfigure}
\begin{subfigure}{0.24\linewidth}
    \includegraphics[width=\linewidth]{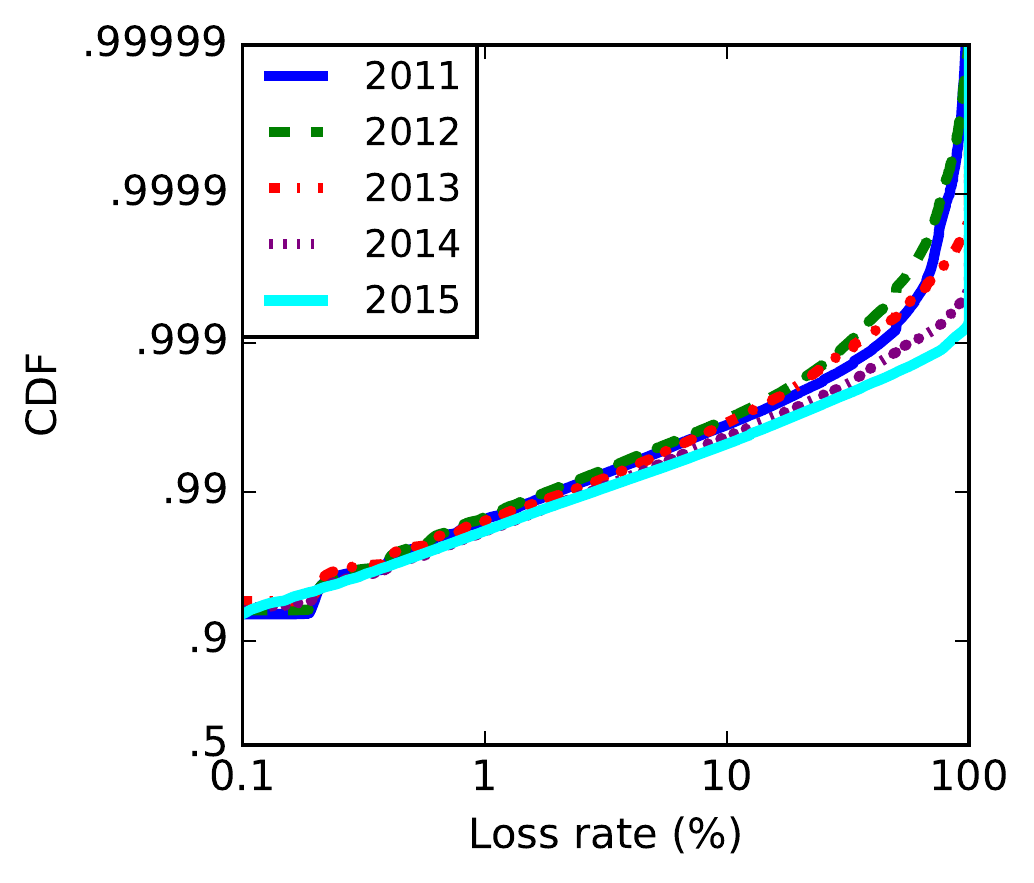}
    \caption{Comcast\label{fig:long-comcast}}
\end{subfigure}
\begin{subfigure}{0.24\linewidth}
    \includegraphics[width=\linewidth]{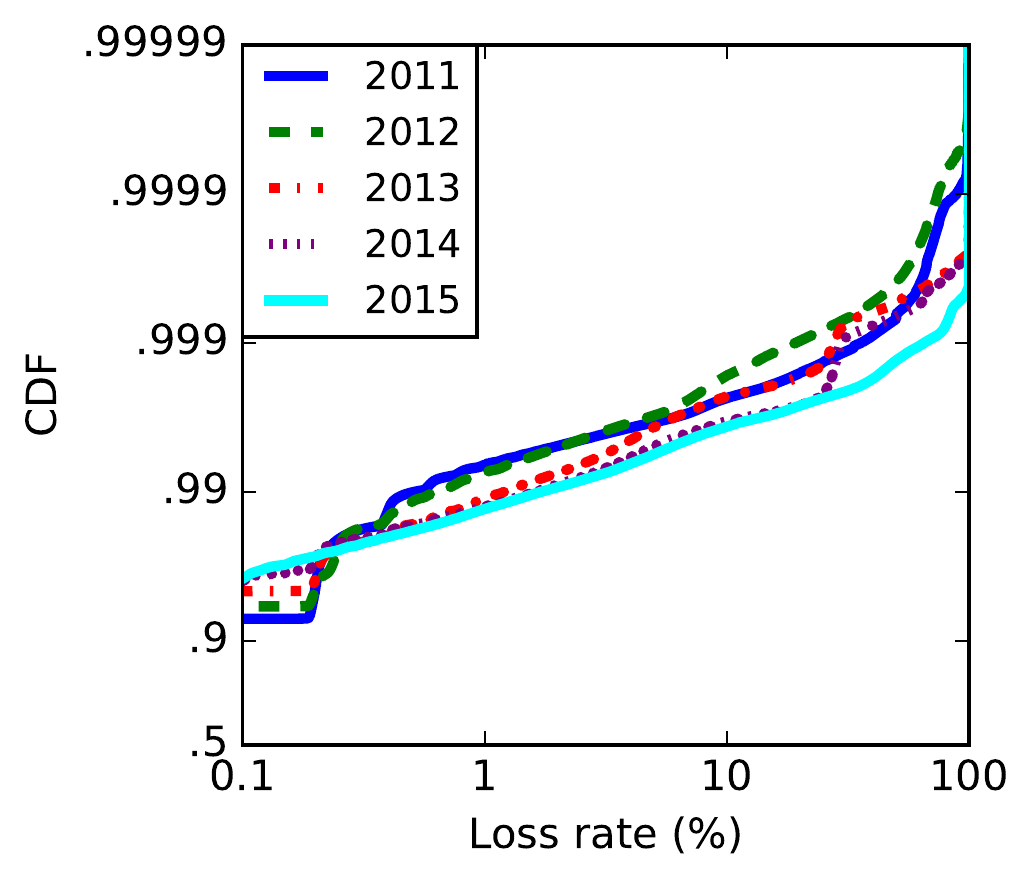}
    \caption{Verizon (fiber)\label{fig:long-verizonfiber}}
\end{subfigure}
\begin{subfigure}{0.24\linewidth}
    \includegraphics[width=\linewidth]{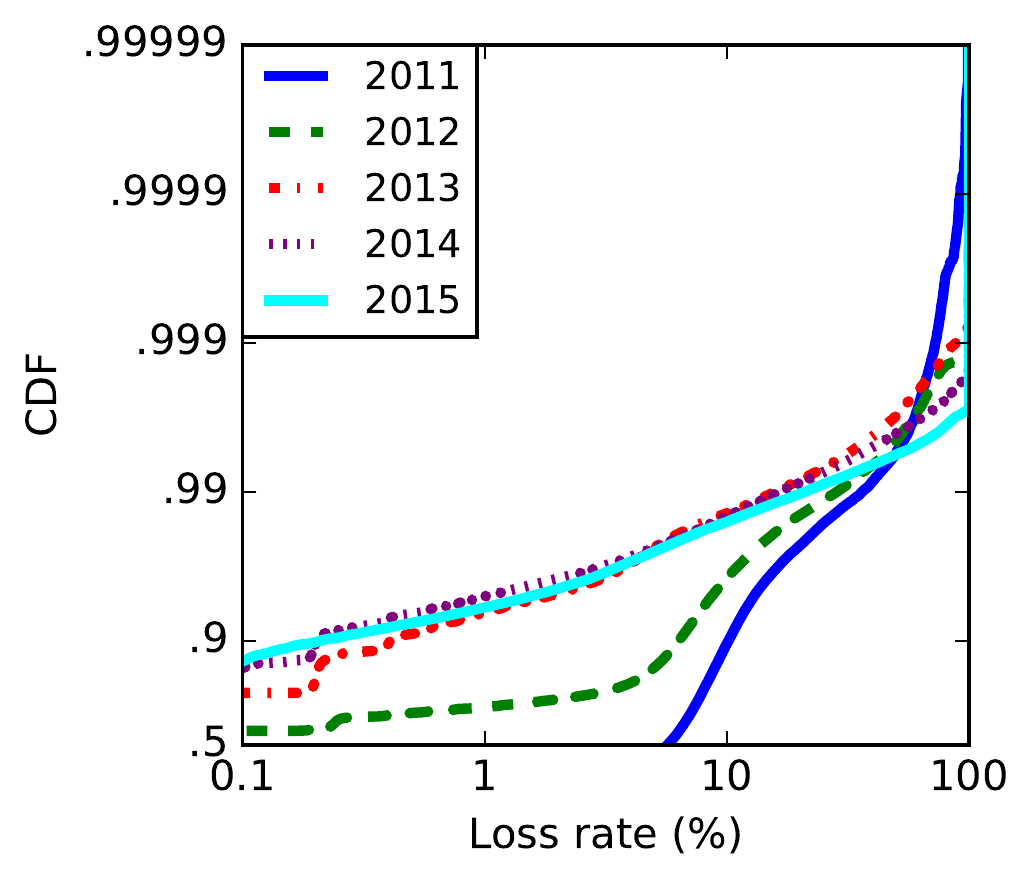}
    \caption{Windblue/ViaSat\label{fig:long-viasat}}
\end{subfigure}
\end{minipage}
\caption{Longitudinal analysis of loss rates for four different ISPs.\label{fig:long}}
\end{figure*}

We close our analysis of broadband reliability with a longitudinal analysis of
ISPs' reliability. With the exception of satellite, which started with very
frequent periods of high loss, service reliability has remained more or less
stable for  most providers over the years.

Figure~\ref{fig:long} shows the longitudinal trends for four ISPs in our
dataset.\footnote{At the time of submission, only the first 8 months of 2016
are currently available on the FCC's website. The full year will likely be
available by the camera ready date.} Though there were some year-to-year
fluctuations, we did not find any consistent trends in terms of reliability
over the course of the study. 

AT\&T showed some of the widest variations for DSL providers across multiple
years. Other providers, such as CenturyLink, Qwest, and Verizon (DSL) tended to
be more consistent below the 10\% loss rate threshold. 

Cable providers tended to be the most consistent. Comcast, shown in
Figure~\ref{fig:long-comcast}, showed some of the most consistency. Other cable
providers such as Cablevision, Charter, and Cox were similar, though some did
have a one year during which it recorded a slightly higher frequency of high
loss rates.

The fiber services, including Verizon's fiber service (shown in
Figure~\ref{fig:long-verizonfiber}), tended to be consistently more reliable than
most other providers using other technologies, but did show some year-to-year
variations. That said, there did not appear to be a trend (i.e., services were
not getting consistently more or less reliable).

Both satellite providers in our dataset did tend to get better over time.
Figure~\ref{fig:long-viasat} shows the annual trend for Viasat. After having
issues in 2011 and 2012, service reliability becomes much more consistent.

With our increasing reliance on broadband connectivity, the reliability of
broadband services has remained largely stable. This highlights the importance
for studies such as this and the need for techniques to improve service
reliability.

\section{Improving Reliability}
\label{sec:multihoming}

Our characterization of broadband reliability has shown that even with a
conservative definition of failure based on sustained periods of 10\% packet loss,
current broadband
services achieve an availability no higher than two nines on average,
with an average downtime of 17.8~hours per year. Defining availability to be
less than 1\% packet loss (beyond which many applications become unusable)
leaves only a single provider of the 19 ISPs in the FCC dataset with two nines
of availability.

Motivated by our findings of both poor reliability and the effect that this unreliabilty
has on user engagement, we aim to improve service
reliability by two orders of magnitude. This will bring broadband reliability to the minimum 
four nines required by the FCC for the public switched telephone service. Our solution 
should {\em improve resilience at the network level}, be 
{\em easy to deploy} and {\em transparent to the end user}.

\begin{itemize-s}
\item {\bf Easy to deploy:} The solution must be low-cost, requiring no significant new 
        infrastructure and the ability to work despite the diversity of devices and home network 
        configurations. It should, ideally, be plug-and-play, requiring little to no manual configuration. 

\item {\bf Transparent to the end user:} The solution should transparently improve reliability, ``stepping in'' 
    only during service interruption.  This transition should be seamless and not require any
    action from the user.

\item {\bf Improve resilience at the network level:} There have been proposals for improving the access 
    reliability within particular applications, such as Web and DNS (e.g.,~\cite{monet,park:codns}). A single,
    network-level solution could improve reliability for all applications. 
        
\end{itemize-s}
%
%
\noindent
Towards this goal, we present a multihoming-based approach for improving broadband
reliability
that meets these requirements. Multihoming has become a viable option for many
subscribers. The ubiquity of broadband and wireless access points and the increased 
performance of cell networks means that many subscribers have multiple alternatives
with comparable performance for multihoming. In addition, several off-the-shelf residential
gateways offer the ability to use a USB-connected  modem as a backup link.\footnote
{For example, 
a wireless 3G/4G connection or a second fixed-line modem as in the case of the Asus'
RT-AC68U~\cite{asus:ac68u}.} While the idea of multihoming is not
new~\cite{smith:multihoming,akella:multihoming}, {\em we focus
on measuring 
its potential for improving the reliability of residential broadband.}


We use active measurements from end hosts and the FCC's Measuring Broadband
America dataset to evaluate our design. We find that (1)~the majority of availability
problems
occur between the home gateway and the broadband access proider
(\S\ref{subsec:multihoming_issues}); (2)~multihoming can provide the
additional two nines of availability we seek
(\S\ref{subsec:multihoming_potential}); and (3)~multihoming to wireless access points
from neighboring residences can often dramatically improve reliability, even when
the neighboring access point is multihomed to the same broadband access ISP
(\S\ref{subsec:multihoming_neighbor}).

\subsection{Where failures occur}
\label{subsec:multihoming_issues}

\begin{table}[t]
\small
\centering
\begin{tabular}{| l | c |}
\hline
Farthest reachable point in network  &  Percent of failures \\  \hline
(1) Reached LAN gateway         & 68\% \\ 
(2) Reached provider's network  & 8\% \\
(3) Left provider's network    & 24\% \\ \hline
\end{tabular}
    \caption{Farthest reachable point in network
        during a connectivity issue, according to 
        traceroute measurements.}
\label{fig:network_issue_type}
\end{table}

We first study where the majority of broadband connectivity issues appear.  We
deployed a network experiment to approximately 6,000 endhosts running
Namehelp~\cite{otto:imc12:dnscdn} in November and December 2014.  For each end
host, our experiment ran two network measurements, a ping and a DNS query, at
30-second intervals. We chose to target our measurements to Google's public DNS
service (i.e., 8.8.4.4 and 8.8.8.8).  For this experiment, we considered this
to be a sufficient test of Internet connectivity.

If neither ping nor a DNS query received a response, we immediately
launched a traceroute to the target.  If the traceroute did not receive a
response from the destination, our experiment recorded the loss of connectivity
and reported the traceroute results once Internet access had been restored.
As in previous work~\cite{feamster:sigmetrics2003}, we used this traceroute data
to categorize the issue according to how far into
the network the traceroute's probes reached.  
Table~\ref{fig:network_issue_type} lists the farthest reachable point in the 
network during a connectivity interruption. 

\textit{We find that most reliability problems occur between the home gateway and the
service provider.  During 68\% of issues, our probes were able to reach the
gateway, but not the provider's network.} We cannot determine whether there was
a problem with the access link, the subscriber's modem, or the gateway configuration, 
but in each case, we ensure that nothing had changed with
the client's local network configuration (e.g., connected to the same access
point and has the same local IP address) and that the probes from the client
reached the target server during the previous test.  Another 8\% of traces were
able to reach the provider's network, but were unable to reach a network beyond
the provider's.  The remaining 24\% left the provider's network, but could not
reach the destination server.

\begin{figure}
    \centering
\begin{subfigure}{0.65\linewidth}
    \centering
    \includegraphics[width=\linewidth]{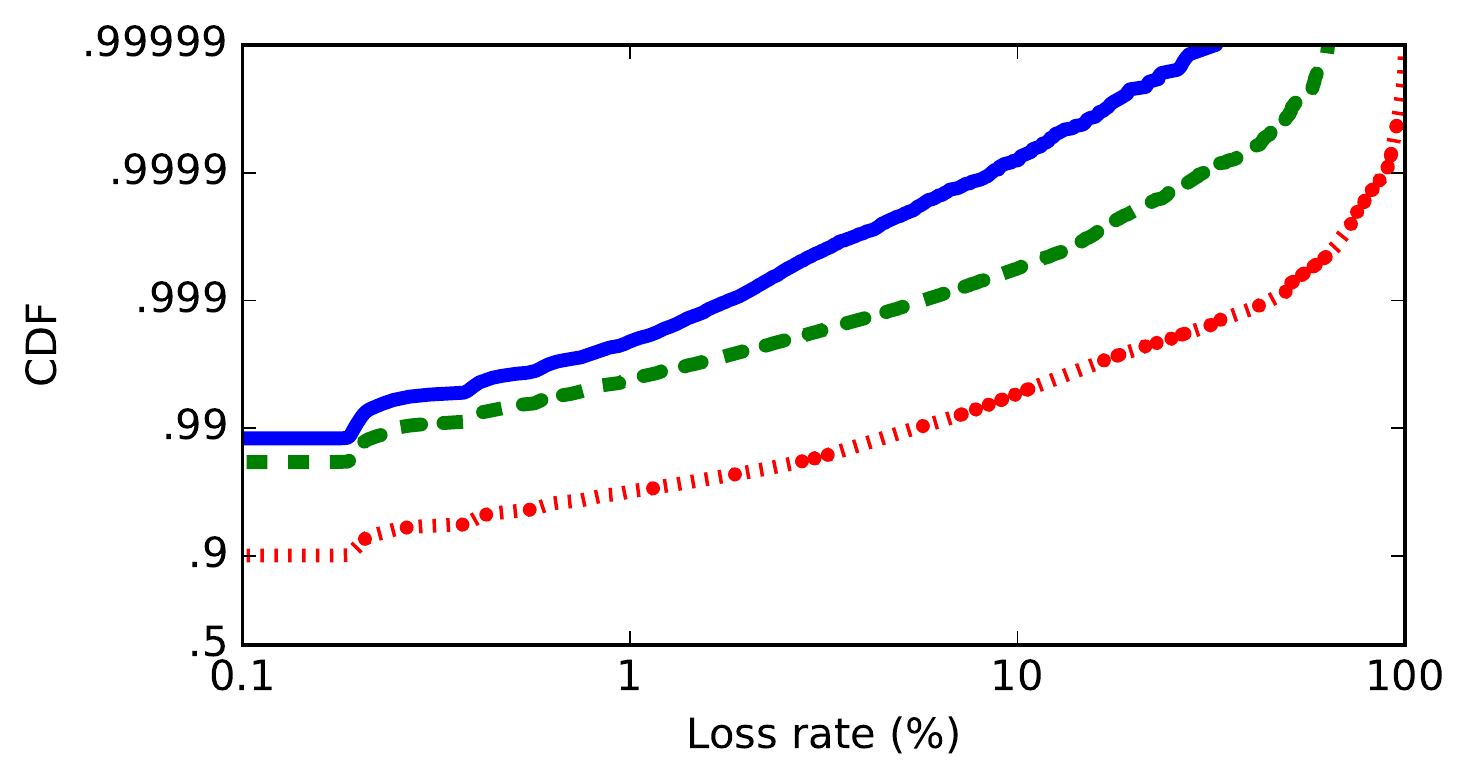}
    \caption{Hourly loss rates
    \label{fig:multihome_loss_cdf}}
\end{subfigure} \begin{subfigure}{0.25\linewidth}
    \centering
\includegraphics[width=\linewidth]{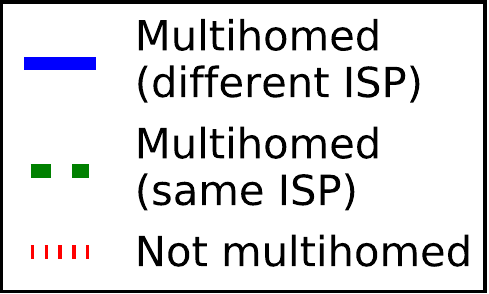}
\end{subfigure}\\
\begin{subfigure}{0.95\linewidth}
    \centering
\includegraphics[width=0.9\linewidth]{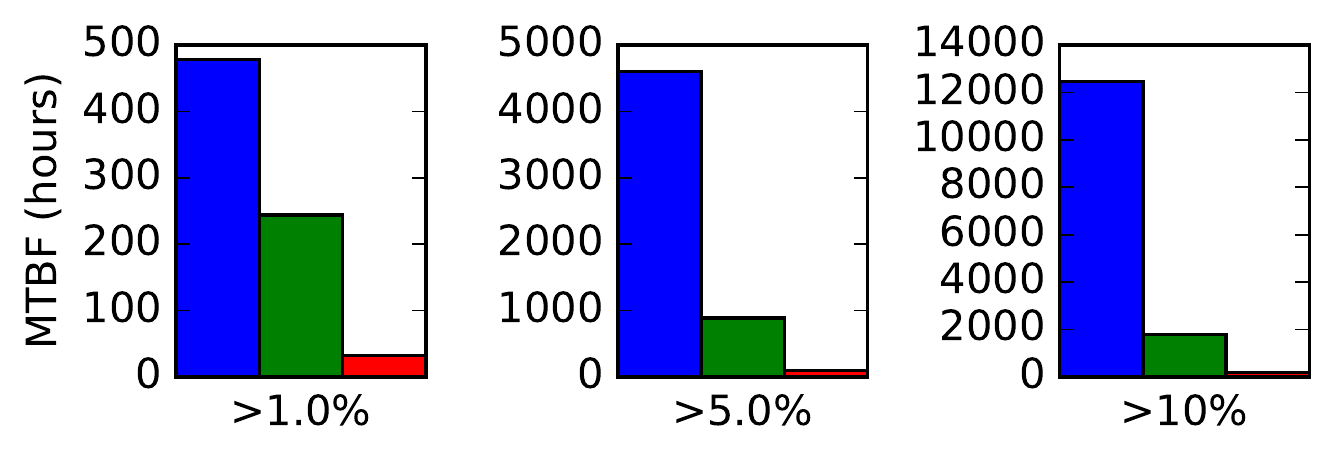}
\caption{MTBF
    \label{fig:multihome_loss_interval_bar}}
\end{subfigure}
\caption{Hourly loss rates, and MTBF measured from each gateway and 
simulated multihomed connection.}
\end{figure}

\subsection{Broadband multihoming}
\label{subsec:multihoming_potential}

Because the majority of service interruptions occur between the home gateway
and the service
provider, we posit that a second, backup connection---multihoming---could improve 
service availability. 



To estimate the potential benefits of broadband multihoming for improving
service reliability, we use the FCC dataset and group study participants by
geographic region based on their Census Bureau block group. A Census block is
the smallest geographical unit for which the Census Bureau publishes data, such
as socioeconomic and housing details. Blocks are relatively small, typically
containing between 600 and 3,000
people.\footnote{\url{https://www.census.gov/geo/reference/gtc/gtc_bg.html}} 
Unfortunately, we are not able to study trends at a finer granularity. 

We identify blocks with at least two users online during the same time period.
For each pair of users concurrently online in a region, we simulate a
multihomed connection by identifying the minimum loss rate between the two
connections during all overlapping time windows.  We distinguished between
simulated multihomed connections depending on whether both users subscribed to
the same ISP.

Figure~\ref{fig:multihome_loss_cdf} shows the results of this experiment as a
CDF of the loss rates reported for each simulated multihomed connection. As a
baseline for comparison, we include the original reported loss rates for the
same population of users, labeled ``Not multihomed''. For both types of simulated
multihomed connections (same and different ISP), high packet loss rates are at
least an order of magnitude less frequent.  Furthermore, the benefits of
multihoming with different ISPs as opposed to using the same ISP increase as
the loss rate threshold increases. For example, using a 1\% threshold as a
failure, both scenarios provide two nines of reliability (99.59\% when using
the same ISP, 99.79\% when using different ISPs).  However, at 10\% loss,
multihoming on the same ISP provides only three nines (99.94\%), while
multihoming on different ISPs provides four nines (99.992\%).

Figure~\ref{fig:multihome_loss_interval_bar} shows the average interval between
periods of high packet loss rates, with thresholds of 1\%, 5\%, and 10\%. Although
both types of multihomed connections improve availability, as the loss rate threshold
increases, the difference between connections multihomed on the same ISP and
connections multihomed on different ISPs increases: with a 10\% packet
loss rate threshold, a multihomed connection using different ISPs provides four
nines of availability, versus three nines for a connection multihomed on the
same provider, and about two nines on a single connection.

\subsection{Neighboring networks to multihome}
\label{subsec:multihoming_neighbor}


There are multiple ways that broadband subscribers could multihome their
Internet connection. One possibility would be for users to subscribe to a
cellular network service, adding to their existing wireless plan. This approach
would be straightforward to implement, as users would only need to add a 4G
dongle to their device. However, the relatively high cost per GB of traffic
would likely be too expensive for most users, preventing them from using
network-intensive services, such as video streaming.

An alternative, and cheaper, realization of our approach could adopt a cooperative
model for multihoming between neighbors either through a volunteer
model~\cite{www:open_wireless,www:fon} or a provider's supported community
WiFi~\cite{pam:xfinity}.\footnote{Providers offering such services include
AT\&T, Comcast, Time Warner, British Telecom (UK) and Orange (France).} 

To show the feasibility of this model, we used Namehelp clients to measure
wireless networks between December 7, 2015 and January 7, 2016. For each user,
every hour we recorded their current wireless configuration and scanned for
additional wireless networks in the area using OS X's \texttt{airport} and
Windows' \texttt{netsh.exe} commands.

One challenge to estimating the number of available APs is that, in many cases,
an individual AP device will host multiple networks (e.g., 2.4 Ghz, 5 Ghz,
and/or guest networks) using similar MAC addresses. To avoid overestimating the
number of available APs, we used multiple techniques to group common SSIDs that
appeared in our wireless scans. We first grouped MAC addresses that were
similar to each other (i.e., string comparisons showed they differed in four or
fewer hexadecimal digits or only differed in the 24 least significant bits).
We then manually inspected these groups and removed any with an SSID that
clearly did not correspond to a gateway, such as network devices and WiFi range
extenders (e.g., SSIDs that contained ``HP-Print'', ``Chromecast'', or
``EXT''). We consider the AP groups remaining as gateway devices.

\begin{figure}[t]
    \centering
\begin{minipage}{1\linewidth}
\begin{subfigure}[t]{0.48\linewidth}
    \centering
\includegraphics[width=\linewidth]{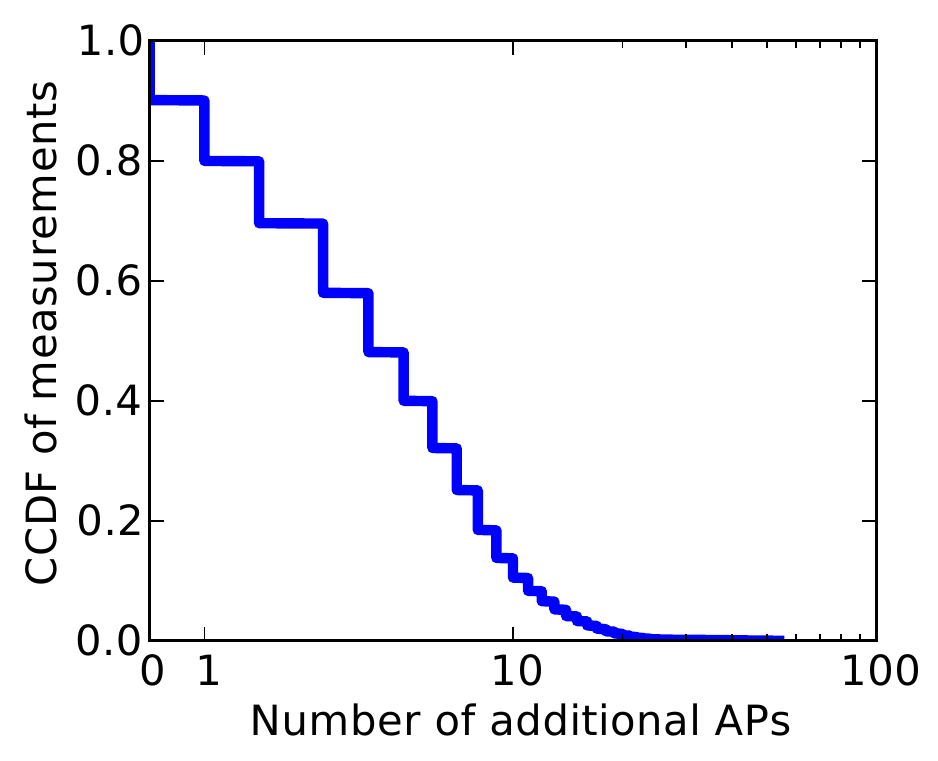}
    \caption{\label{fig:ccdf_unique_networks}}
\end{subfigure} 
\begin{subfigure}[t]{0.48\linewidth}
    \centering
\includegraphics[width=\linewidth]{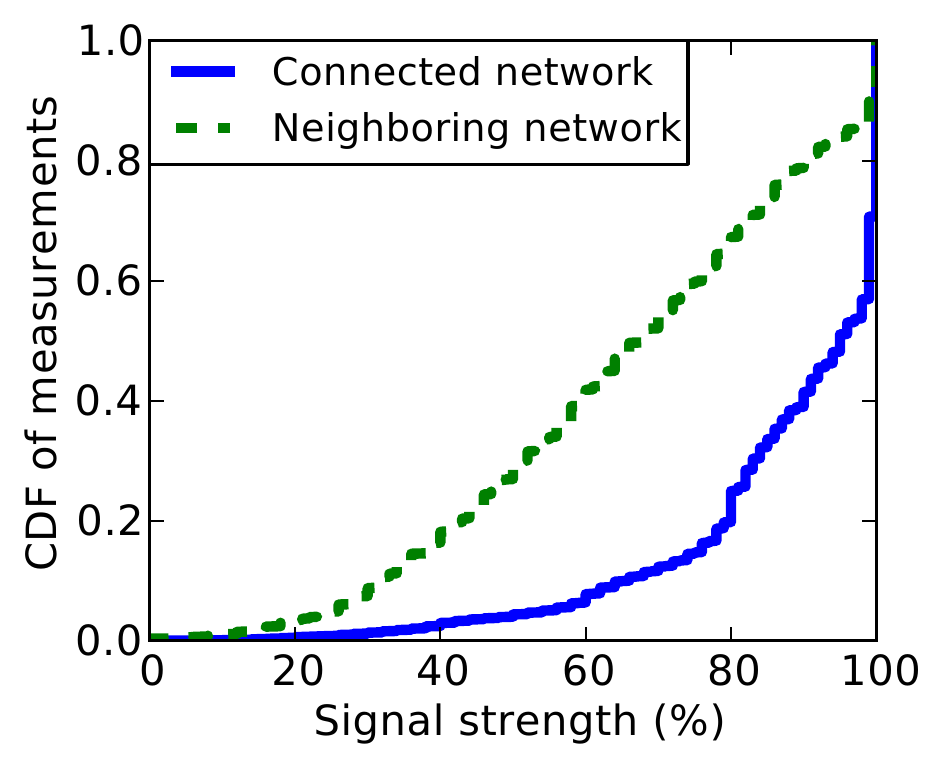}
\caption{\label{fig:sig_str_cdf}}
\end{subfigure}
\end{minipage}
\vspace{-8pt}
\caption{Number of additional APs available (a) and signal strengths for the
current and strongest alternative AP (b).}
\vspace{-8pt}
\end{figure}

Figure~\ref{fig:ccdf_unique_networks} shows the CCDF of the number of
additional unique groups seen across all measurements. Since we combine 
our findings at the end of this section with those of the previous section 
(\S\ref{subsec:multihoming_potential}), we only include measurements collected 
from clients within the US in our analysis. In 90.2\% of cases, one
or more {\em additional} wireless APs are available to the client. In
approximately 80\% of
cases, two or more additional APs are available.  These results highlight the
potential for using nearby APs to improve service availability via multihoming.

The availability of neighboring AP is a necessary but not sufficient condition; a 
remaining concern is whether clients would actually be able to connect to
these APs. Figure~\ref{fig:sig_str_cdf} shows a CDF of the signal
strength percentage of both the AP to which the client is currently connected
as well as the signal of the strongest available alternative network
(``Neighboring network'').  While the signal strengths of the neighboring
networks are typically lower than that of the home network, it is still
sufficiently strong in most cases, with a signal strength of 40\% or higher for
82.7\% of measurements. 

Last, to estimate the potential improvement in service availability of 
using a neighboring AP as a backup connection, we infer the ISP of an AP by analyzing
its SSIDs. For example, we found a large number of APs advertising SSIDs that
clearly identify the provider, such as those starting with ``ATT'' and
``CenturyLink''.  Similarly, we classified APs that hosted an ``xfinitywifi''
network in addition to other SSIDs as neighboring networks that belonged to
Comcast subscribers. We were able to infer the ISP of at least one neighboring
AP in 45\% of all scans. Of these, 71\% of APs appeared to belong to
subscribers of an ISP different from that of the client. 

In conjunction with the results from Section~\ref{subsec:multihoming_potential},
these findings
suggest that if clients used these additional APs for backup connections, service
availability
would improve by two nines in at least 32\% of cases and by one nine in at least an additional
13\% of cases.  Since many APs advertised user-defined or manufacturer default 
SSIDs (e.g., ``NETGEAR'' or ``linksys''), this is a lower bound estimate of the potential of 
improving service availability through multihoming. The next section presents a
prototype system, \system{}, that is based on this insight.

\section{AlwaysOn}
\label{subsec:multihoming_system}

In this section, we discuss various challenges associated with broadband multihoming;
we describe how we address these concerns in our prototype
service, {\em AlwaysOn}; and evaluate its performance.

\begin{figure}[t]
    \centering
\includegraphics[width=0.85\linewidth]{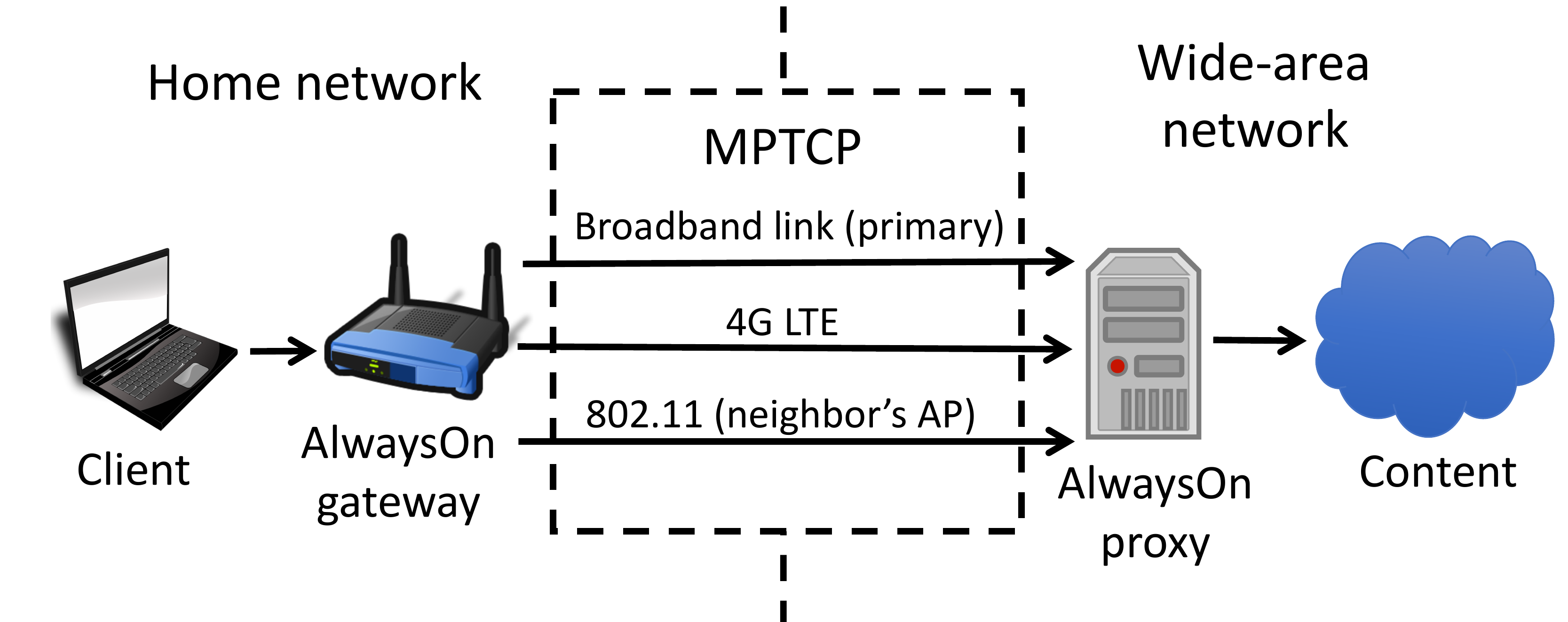}
\vspace{-8pt}
\caption{Two {\em AlwaysOn} configurations using a neighboring AP or a 4G hotspot.  
The black solid line represents a client's normal path while the gray lines
represent possible backup routes.
    \label{fig:mptcp_config}}
\vspace{-8pt}
\end{figure}


\begin{figure}[t]
    \centering
\begin{subfigure}{0.48\linewidth}
    \centering
    \includegraphics[width=\linewidth]{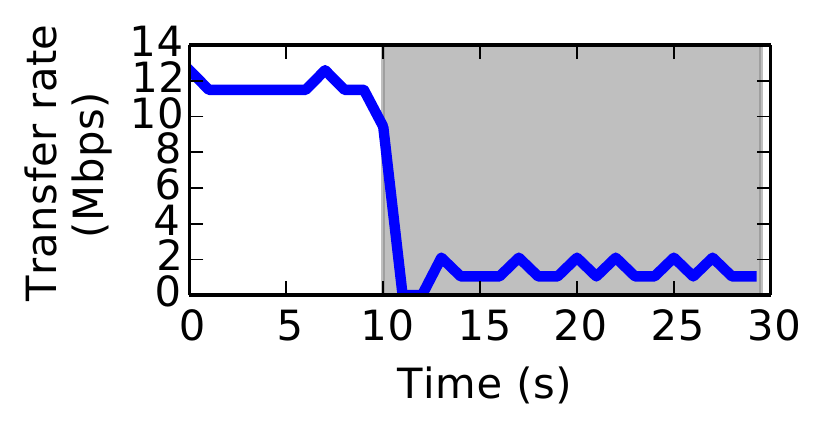}
    \caption{Comcast / ATT \\
                (75 Mbps / 3 Mbps) \label{fig:multi_comcast-att}}
\end{subfigure} 
\begin{subfigure}{0.48\linewidth}
    \centering
    \includegraphics[width=\linewidth]{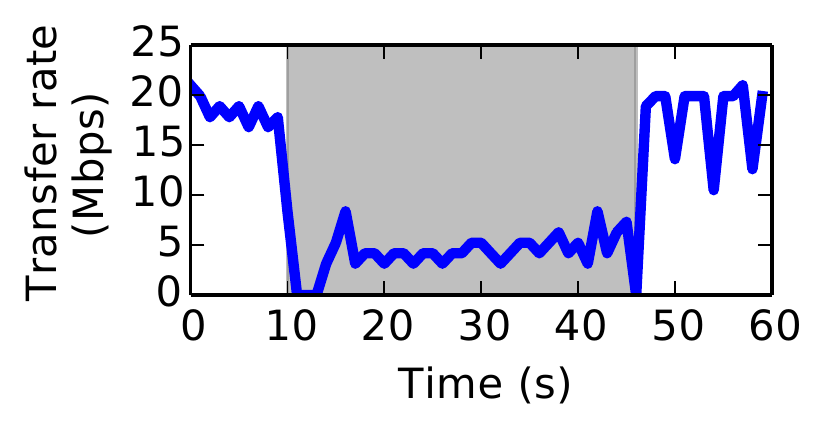}
    \caption{RCN / VZW \\
                (150 Mbps / 4G LTE) \label{fig:multi_rcn-verizon}}
\end{subfigure}
\vspace{-8pt}
\caption{Throughput using {\tt iperf} using {\em AlwaysOn} in two different 
network deployments. Each figure lists the service providers and speeds for
the primary and
secondary connections. The gray shaded section of the graph timeline represents
the time during which the we simulated an outage on the primary 
link.\label{fig:multi_all}}
\end{figure}




\subsection{Design Challenges}

Multihoming a residential broadband connection presents different challenges
than conventional multihoming. Whether failing over to a neighbor's wireless
access point or to a 4G connection, a na\"{i}ve implementation may interrupt
the clients' current open connections and require re-opening others,
because switching connections will cause the client to have a different source
IP address for outgoing traffic. A broadband multihoming solution should be
able to seamlessly switch between the primary and secondary connections
without interrupting the user's open connections.

Broadband multihoming also introduces concerns related to usage policies and
user privacy. First, some backup connections  (e.g., 4G) may have data caps. A
common broadband use like streaming ultra  HD videos may have to be restricted
over those connections considering their cost. Neighbors sharing connections
with each other may also prefer to impose limits on how they share their
connection, for instance, in terms of available bandwidth, time or total
traffic volume per month. Second, in locations where there is more than one
alternative backup connection, users may want to state their preference in the
order of which networks to use based on factors such as the structure of their
sharing agreement or the amount of wireless interference on a network's
frequency.  Finally, there are privacy concerns for both parties when
multihoming using a neighbor's network. Users that ``borrow'' their neighbor's
network would, by default, be allowing their neighbor to capture their
unencrypted traffic. Conversely, neighbors who are ``loaning'' access to their
network should not have to compromise their own privacy to do so.

\subsection{AlwaysOn Design \& Implementation}


AlwaysOn has two components: a client running in the gateway
and a proxy server deployed as a cloud service. A diagram of this deployment 
is shown in Figure~\ref{fig:mptcp_config}. The additional lines in this figure 
represent a backup paths via a neighboring AP and another through a 4G hotspot. 

AlwaysOn uses Multipath TCP (MPTCP)\cite{mptcp} to transition connections from
primary to secondary links without interrupting clients' open connections. The
AlwaysOn Gateway  creates an encrypted tunnel to the MPTCP-enabled AlwaysOn
Proxy. All traffic from the private LAN is routed via this tunnel. Our current
implementation uses an encrypted SOCKS proxy via SSH; a virtual private network
(VPN) tunnel would be an alternate implementation. Using an encrypted tunnel
ensures that user traffic remains confidential when it traverses the backup
path. The MPTCP-enabled proxy can be shared between multiple users, with each
user being assigned (after authentication) to a unique tunnel.

Additionally, the AlwaysOn gateway sends the traffic via a guest network that
is isolated from the private LAN when sharing its connection; many commodity
residential access points already offer guest network configuration to enable
connection sharing while limiting a guest's access to the local network.
Deploying AlwaysOn requires an MPTCP-enabled kernel; although home network
testbed deployments (e.g., BISmark) do not run such a kernel, OpenWrt can be
built with mptcp support. 

AlwaysOn allows users to configure traffic shaping settings to facilitate
resource sharing across connections. Options that concern traffic
shaping must be synchronized between the gateway and proxy. The gateway can
shape outgoing traffic, but incoming traffic must be shaped on the AlwaysOn
Proxy. Our current prototype uses \texttt{tc} and \texttt{iptables} to enforce
traffic management policies. For outgoing traffic, the AlwaysOn gateway can
throttle traffic traversing  the neighboring access point, as well as traffic
on its own guest network.  Each user has unique port number at the gateway to
use for their tunnel, and the IP address serves to identify traffic to or from
secondary links. Using \texttt{iptables}, the gateway and proxy mark traffic according
to whether it corresponds to a primary or secondary connection. They then
use \texttt{tc} to apply the appropriate traffic shaping policy.

A user must currently manually configure policies such as link preference and
traffic shaping at the AlwaysOn Gateway and proxy; this manual configuration
poses a problem when a user needs to configure settings such as link
preference on devices that they may not necessarily control. We are exploring 
alternatives to realize this through a third-party service that
accepts, encodes, and enforces such policies on outgoing and incoming traffic.

\subsection{AlwaysOn Evaluation}

We ran multiple experiments to evaluate the AlwaysOn prototype in
various network settings.  We instantiated an AlwaysOn
proxy server on a university network. We aim to evaluate 
AlwaysOn in two operating modes: (1)~during the failure of the primary link; and
(2)~during normal operation.  Routing traffic
via the AlwaysOn proxy should not affect performance during normal
operation, considering that even the least reliable service still has about
36 hours of downtime each month. In addition, to limit the effect of outages on
user quality of experience, AlwaysOn should respond quickly and route traffic via the backup
link as soon as a failure is detected.

\paragraph{Reaction to network failures.}
In our first experiment, we test AlwaysOn's ability to react to network
failures. We ran {\tt iperf} for 30 seconds from a client behind the AlwaysOn
Gateway, recording {\tt iperf}'s measured throughput rate each second. We
emulated different outages, represented in the plots by time periods
highlighted in gray. 

We ran this experiment in two different scenarios for our evaluation, shown in
Figure~\ref{fig:multi_all}.  
In the first scenario (Figure~\ref{fig:multi_comcast-att}), we used a Comcast
75~Mbps service as the primary connection and a 3~Mbps AT\&T service as the
secondary connection. In the second scenario, shown in
Figure~\ref{fig:multi_rcn-verizon}, we used an RCN 150~Mbps service as the
primary, and a Verizon Wireless 4G LTE hotspot as the secondary connection. For
this test, the primary connection was re-enabled after approximately
35~seconds. Once the primary connection was reestablished, AlwaysOn switches
traffic back to the RCN connection.

In each case, AlwaysOn can recover relatively quickly once it realizes the
primary link is no longer working, and does not require the connection to be
reestablished.  We also ran {\tt iperf} over each connection between the same
client and server without the AlwaysOn gateway and proxy and consistently
measured similar throughput rates.  The relatively slow performance compared to
the access link speed in 
Figures~\ref{fig:multi_comcast-att} and~\ref{fig:multi_rcn-verizon} is likely
due to other limiting factors such as end-to-end latency, congestion on the
path, and only using a single TCP connection.

\paragraph{Application performance during access network failures.} We also
evaluated how AlwaysOn was able to handle service outages while clients used an
OTT service. We ran tests using both Netflix and HTTP Live Streaming (HLS)
services. For each service, we tested how outages affected playback of the
stream for a non-multihomed and multihomed configurations (using AlwaysOn and
two 150~Mbps RCN connections).

\begin{figure}[t]
    \centering
\includegraphics[width=.88\linewidth]{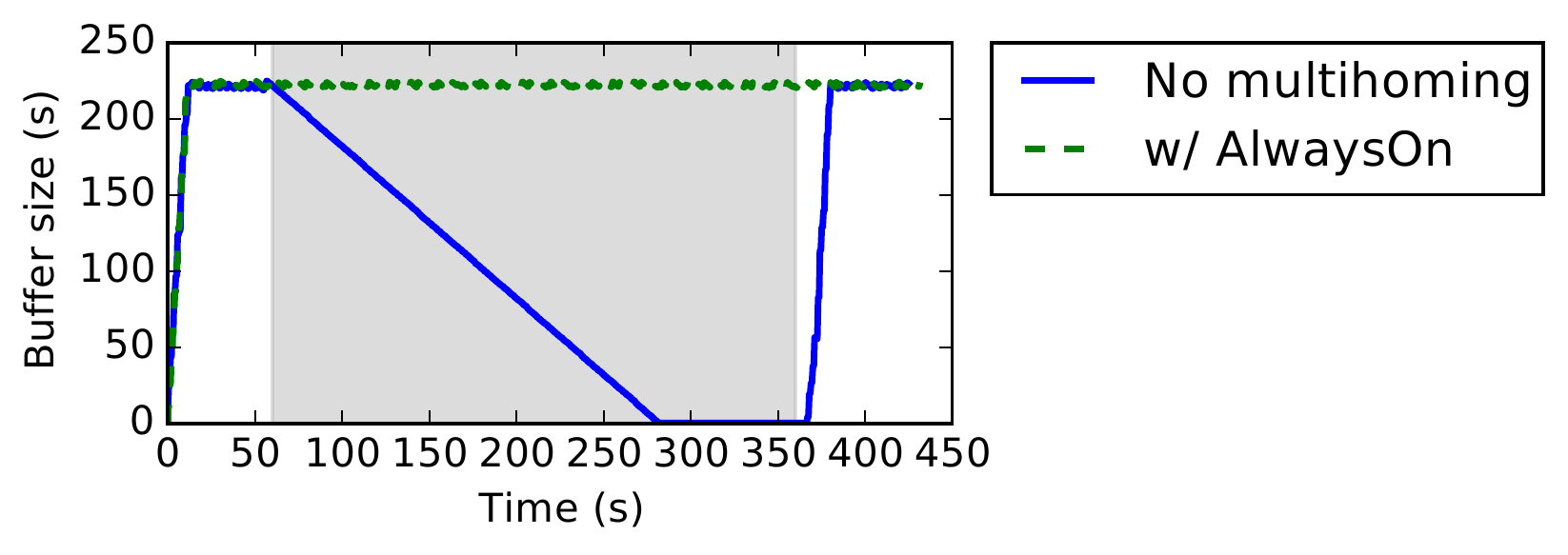}
\vspace{-10pt}
\caption{Two Netflix video streaming sessions, one without any multihoming 
and one while using AlwaysOn. The gray highlighted section represent
simulated network outages on the primary link. \label{fig:netflix}}
\vspace{-12pt}
\end{figure}

Figure~\ref{fig:netflix} shows the playback of a Netflix stream for both
configurations. In this test, we streamed a video for one minute, allowing the
stream's buffer to stabilize at $\approx$220~seconds. We then simulated a five
minute outage on the primary link. Without multihoming, any outage lasting
longer than about 3.5 minutes resulted in the buffer being drained completely,
interrupting playback of the stream. Once the connection is restored, the
non-multihomed host is able to quickly resume the stream. In the experiment
using AlwaysOn, the stream is uninterrupted as the traffic was able to
seamlessly be routed via the secondary connection.

\begin{figure}[t]
    \centering
\begin{subfigure}{0.46\linewidth}
    \centering
    \includegraphics[width=\linewidth]{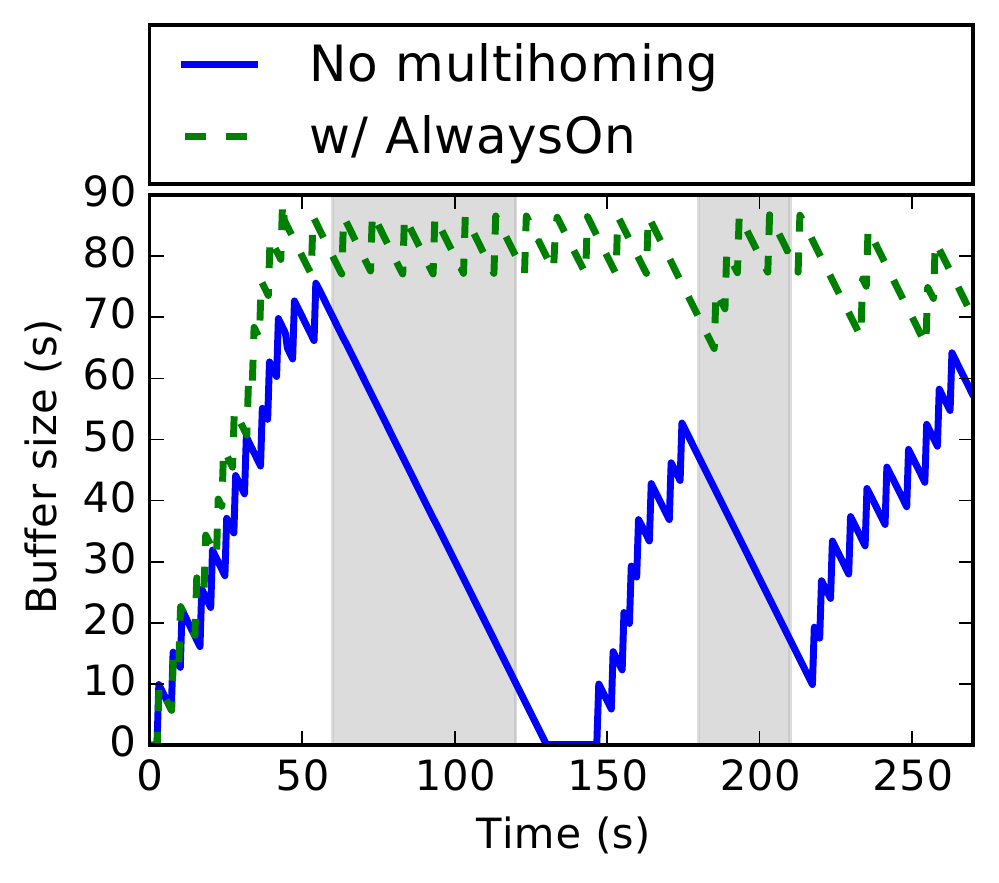}
    \caption{Buffer size\label{fig:hls_buffer}}
\end{subfigure}
\begin{subfigure}{0.49\linewidth}
    \centering
    \includegraphics[width=\linewidth]{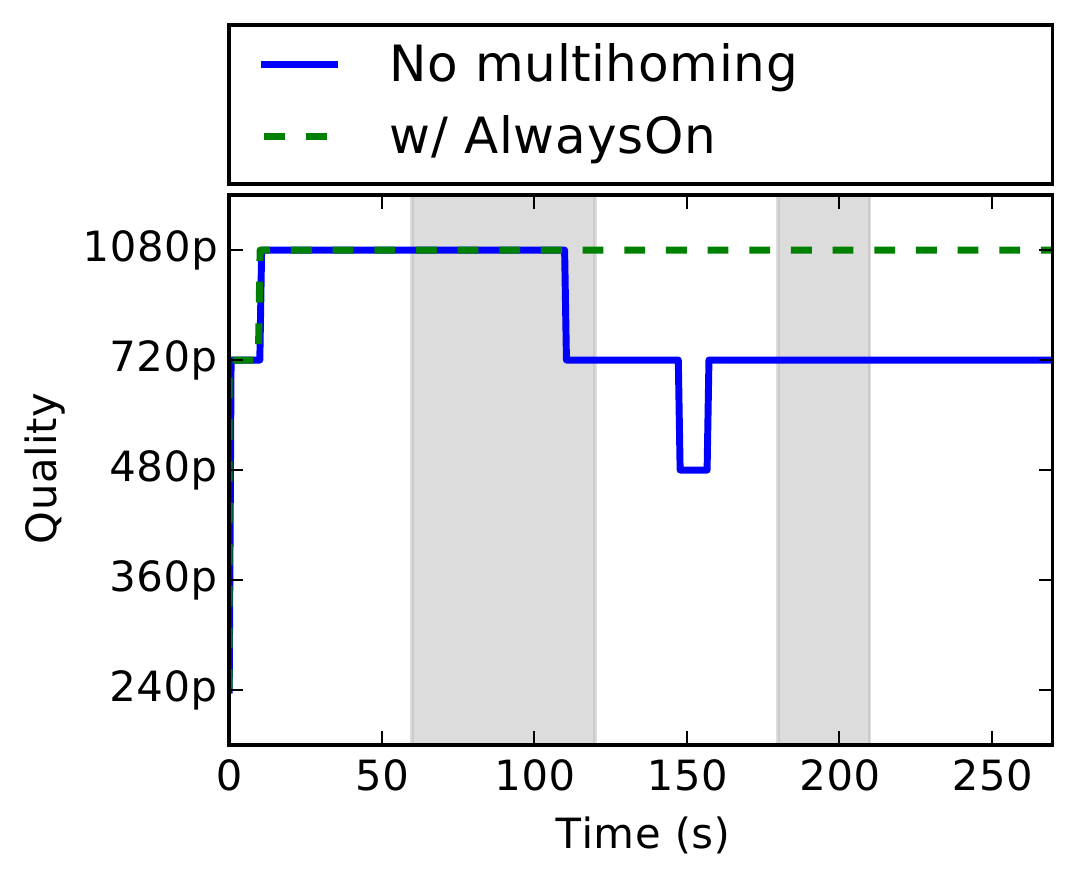}
    \caption{Video quality \label{fig:hls_quality}}
\end{subfigure}
\vspace{-8pt}
\caption{Comparison of two HLS video sessions, one without multihoming 
and one while using AlwaysOn. The gray highlighted sections represent
simulated network outages on the primary link. \label{fig:hls_compare}}
\vspace{-12pt}
\end{figure}

Figure~\ref{fig:hls_compare} shows the result of this experiment using HLS. In
this test, we evaluated the impact of shorter outages.
Figure~\ref{fig:hls_buffer} shows the size of the buffer;
Figure~\ref{fig:hls_quality} shows the current video quality being displayed to
the user.  In both configurations, we simulated two shorter outages that were
one minute apart, the first lasting for 60~seconds and the second lasting for
30~seconds.

In both scenarios, the buffer is able to fill quickly during the initial 60
seconds of the stream, with playback quality quickly increasing to 1080p.
Without AlwaysOn, the buffer is completely drained during the first outage and
the stream is interrupted. It then resumes the stream at 480p for a short
period. The second outage then forces the stream to stay at 720p for the
remainder of the stream. In contrast, the AlwaysOn configuration 
maintains quality at 1080p shortly after initializing the stream, even during the
outage period.

\begin{figure}
    \centering
\begin{minipage}{1\linewidth}
    \centering
\begin{subfigure}{0.49\linewidth}
    \centering
\includegraphics[width=\linewidth]{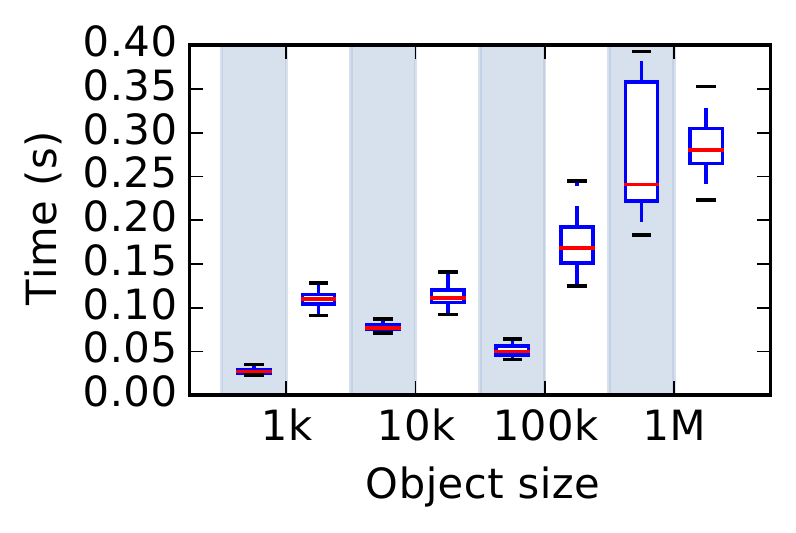}
    \caption{RCN \\ (150 Mbps)
            \label{fig:multi_perf_rcn}}
\end{subfigure} 
\begin{subfigure}{0.49\linewidth}
    \centering
\includegraphics[width=\linewidth]{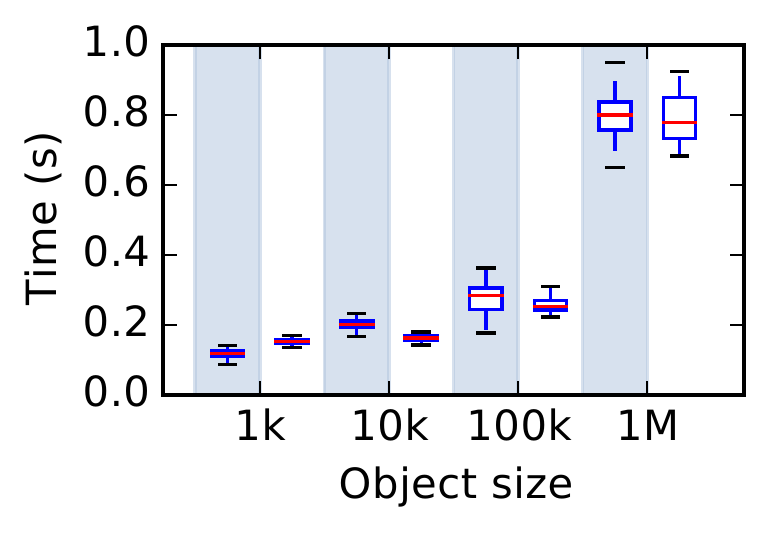}
    \caption{VZW \\ (4G LTE)
            \label{fig:multi_perf_verizon}}
\end{subfigure} 
\end{minipage}
\vspace{-8pt}
\caption{Box and whisker plots showing the time to fetch objects hosted by
Akamai while using (highlighted) and not using the AlwaysOn proxy.
for RCN and Verizon Wireless (VZW).  Each box and whisker represents the median, 
interquartile range (IQR), and 1.5 IQR for each experiment configuration.\label{fig:multi_perf_all}}
\vspace{-12pt}
\end{figure}

\paragraph{Performance overhead of AlwaysOn.}
To see how our AlwaysOn proxy affected network performance, we also measured
the time to fetch objects hosted on Akamai's CDN, both when using and not using
the proxy.  For this test, we downloaded files of varying sizes (1~kB, 10~kB,
100~kB, and 1~MB) 100 times. The box plots shown in
Figure~\ref{fig:multi_perf_all} summarize the distribution of download times
for objects of each size while using the RCN (Figure~\ref{fig:multi_perf_rcn})
and Verizon Wireless (Figure~\ref{fig:multi_perf_verizon}) connections.  The
highlighted box plots show the fetch times while using the proxy for each
respective file size. 
Clients who forwarded traffic via AlwaysOn experienced similar performance; in some
cases, we found that download times actually improved while using the AlwaysOn 
proxy, since clients were directed to a much faster replica when using the proxy.


%

\section{Related Work} \label{sec:related}
 
Previous work on broadband networks has often focused on characterizing
services in terms of performance (e.g., link capacity and latency) from a range
of platforms and vantage points, including customized home
gateways~\cite{samknows, bismark}, applications in end-user
devices~\cite{bischof:wmust11, dicioccio:homenet, sanchez:nsdi,
otto:imc12:dnscdn}, Web-based tests~\cite{kreibich:netalyzr, canadi:revisiting,
ritacco:hmn}, and well-provisioned measurement nodes outside the access
networks~\cite{dischinger:residential}. We use longitudinal data collected by
two of these efforts, the FCC's MBA initiative and
Namehelp~\cite{otto:imc12:dnscdn}, to study broadband reliability.

\paragraph{Reliability of phone networks.} A number of efforts have tackled
reliability characterization in other contexts, such as telephone and cellular
networks. Thanks in part to the pioneer work at Bell
Labs~\cite{dhillon:reliability} by the end of the 20th century, public switched
telephone networks (PSTN) had become so reliable that AT\&T expected no more
than two hours of failure over a 40-year period~\cite{kuhn:pstn}. Today the FCC
requires that PTSN providers document and report outages affect more than
30,000 users or last longer than 30 minutes~\cite{patterson:lessons} which
corresponds with at least four nines of availability. 

\paragraph{Effect of network factors on user behavior.} Recent work has
explored the effect of network factors on user experience with applications,
including VoIP~\cite{chen:skype}, Web~\cite{balachandran:web-cell-qoe} and
Internet video~\cite{balachandran:video}. Rather than deriving a model for user
experience based on multiple factors, our work focuses on the effect of
reliability on user demand. Others have started to explore the use of
alternative experimental designs to evaluate user experience.  Krishnan et
al.~\cite{sitaraman:qed} apply quasi-experimental design to evaluate the effect
of video stream quality on viewer behavior, Oktay et al.~\cite{oktay:social}
relies on it for causal analysis of user behavior in social media. Bischof 
et al.~\cite{bischof:imc14} explores the effect of contextual factors such 
as price and competition on user demand. We apply similar methods
to understand the effect of service reliability on user behavior. 

\paragraph{Reliability of broadband providers.} Baltr\={u}nas et
al.~\cite{baltrunas:mobile-reliability} presented a study of the reliability of
four mobile broadband providers in Norway using the Nornet Edge dedicated
infrastructure. This work illustrates the value of end-to-end measurements to
identify failures and performance events not always captured by the operators'
monitoring systems.  Broadband reliability has received little attention until
recently. Lehr et al.~\cite{lehr:broadband-reliability} discuss some of the
challenges of characterizing reliability and their economic and policy
implications and identify three different ways in which the ``reliability'' of
broadband services can be measured: (1)~the reliability of the service itself;
(2)~the reliability of network services offered by the ISP (e.g., DNS); and
(3)~the consistency of the service's performance. We focus on characterizing
reliability in terms of the former two categories and leave the latter as
future work.

\paragraph{Multihomed access networks.} Beyond improvements in access link
technology, one way to enhance the reliability of access networks is through
redundancy. Gummadi et al.~\cite{gummadi:sosr} propose a detouring approach to
recover from Internet path failures. Andersen et al. improved web availability
with their system, MONET, an overlay network of multihomed proxy
servers~\cite{monet}. We have seen the recent introduction of consumer-grade
residential gateways that support a second WAN connection (such as a 3G or 4G
modem)~\cite{asus:ac68u} and some work exploring the performance benefits of
the on-loading of broadband traffic using a 3G connection~\cite{rossi:3gol}. 
Detal et al. presented MiMBox, a system for translating between TCP and MPTCP
connections at the middlebox~\cite{multipath:middlebox}. Other works have
explored the benefits of using MPTCP on mobile
devices~\cite{multipath:wireless,multipath:handover} and the possibility of
bonding multiple access links (such as DSL and cable) to increase 
performance~\cite{mptcp:bonding, habib:multihoming,thompson:multihoming}.



\section{Conclusion}
\label{sec:conclusion}

As broadband performance and availability continue to improve and users migrate
services to over-the-top alternatives, reliability will become the dominant feature
in the evaluation of broadband services. 

We empirically demonstrated the importance of broadband reliability on users'
quality  of experience. We presented an approach for broadband reliability
characterization  using data collected by common national efforts to study
broadband and discussed key  findings from applying it to four-year dataset
collected through the FCC's MBA program.  Motivated by our findings on both
the importance of reliability and the current  reliability of broadband
services, we presented the design, implementation, and evaluation of AlwaysOn,
practical approach to improving broadband reliability through multihoming.
There are a number of promising research directions for future work from alternative 
metrics for broadband service reliability to alternative approaches to study the effect 
of reliability on user behavior.
To facilitate reproducibility and use of the system, we have publicly released our
dataset, analysis scripts, and {\em AlwaysOn} prototype~\cite{alwayson}.

\label{lastpage}

\end{sloppypar}



\small
\setlength{\parskip}{-1pt}
\setlength{\itemsep}{-1pt}
\balance\bibliography{reliability}
\bibliographystyle{abbrv}

\end{document}